\documentclass[prl,twocolumn,letterpaper, superscriptaddress,longbibliography]{revtex4-1}
\usepackage{graphicx}
\usepackage{dcolumn}
\usepackage{bm}
\usepackage{color}
\usepackage{tabularx}
\usepackage{array}
\usepackage{amsmath}
\usepackage{color,soul}

\newcommand{\kbf}{{\mathbf{k}}}

\usepackage{pbox}
\usepackage{stmaryrd}  

\makeatletter

\@ifundefined{textcolor}{}
{%
 \definecolor{BLACK}{gray}{0}
 \definecolor{WHITE}{gray}{1}
 \definecolor{RED}{rgb}{1,0,0}
 \definecolor{GREEN}{rgb}{0,1,0}
 \definecolor{BLUE}{rgb}{0,0,1}
 \definecolor{CYAN}{cmyk}{1,0,0,0}
 \definecolor{MAGENTA}{cmyk}{0,1,0,0}
 \definecolor{YELLOW}{cmyk}{0,0,1,0}
}

\makeatother
\begin{document}

\title{Pump-induced magnon anticrossing due to three-magnon splitting and confluence}

\author{Tao Qu}
\thanks{qutao@unc.edu.}
\affiliation{Department of Physics and Astronomy, University of North Carolina at Chapel Hill, Chapel Hill, NC 27599, USA}

\author{Yuzan Xiong}
\affiliation{Department of Physics and Astronomy, University of North Carolina at Chapel Hill, Chapel Hill, NC 27599, USA}

\author{Xufeng Zhang}
\affiliation{Department of Electrical and Computer Engineering, Northeastern University, Boston, MA 02115, USA}
\affiliation{Department of Physics, Northeastern University, Boston, MA 02115, USA}

\author{Yi Li}
\affiliation{Materials Science Division, Argonne National Laboratory, Argonne, IL 60439, USA}

\author{Wei Zhang}
\thanks{zhwei@unc.edu.}
\affiliation{Department of Physics and Astronomy, University of North Carolina at Chapel Hill, Chapel Hill, NC 27599, USA}

\date{\today}

\begin{abstract}
We present a plausible mechanism for achieving magnon-magnon level repulsion spectrum originating from the oscillation between the splitting and confluence in a three-magnon scattering process. When a magnetostatic mode on a YIG sphere is pumped by a microwave signal near the magnon resonance frequency with an increasing amplitude, the generated magnon condensate at half of the pumping frequency exerts a back-action to the original magnon mode. Such a strong nonlinear coupling manifests a striking feature of a `bending effect' of the magnetostatic mode spectra, akin to the anti-crossing observed in a strongly coupled magnon-photon system.

\end{abstract}

\maketitle

\section{Introduction}

Hybrid magnonics is an emerging interdisciplinary field that explores coherent interactions of magnons with other type of excitations such as photons, phonons, spins and magnons themselves \cite{LachanceQuirionAPEx2019,LiJAP2020,Awschalom21}. The mode hybridization is usually achieved by a strong coupling of magnon with another resonant mode, manifesting a mode anticrossing in the frequency spectrum \cite{HueblPRL2013,TabuchiPRL2014,ZhangPRL2014,GoryachevPRApplied2014,BaiPRL2015,LiPRL2019_magnon,HouPRL2019}. The capability of maintaining phase coherence has led to the realization of many novel and interesting physical phenomena with magnons, such as level attraction \cite{HarderPRL2018}, nonreciprocity \cite{WangPRL19,ZhangPRApplied19}, exceptional points \cite{YouNComm17,ZhangPRL19}, and floquet engineering \cite{ZhangPRL20}. They benefit from the high controllability of magnon system, including frequency tunability with magnetic field and coupling control by changing the spatial location \cite{xiong2024combinatorial}. These phenomena open opportunities in magnon-based coherent information processing and offer new potentials in quantum magnonics \cite{LachanceQuirionScience2020,YouPRL23,YuanPhysRep22,flebus20242024}. 

One unique property of magnons is their nonlinearity, which is at the heart of magnonics both from the fundamental understanding and the technological leverage \cite{YuanJAP23,makiuchi2024persistent}. When the population of a magnon eigenmode is excited above a threshold value, the mode will start to overlap with other eigenmodes, leading to a finite coupling due to magnetic dipolar and exchange interactions \textcolor{black}{\cite{PattonPRB03,wojewoda2023observing,PRL207203,korber2023pattern}}. This amplitude-dependent nonlinear coupling leads to energy redistribution among different magnon eigenmodes by three-magnon or four-magnon scattering \cite{PattenPRB09,Kurebayashi11,LiuPRB19,SchultheissPRL09}, which is referred to as the Suhl instability\cite{Suhl57}. The process usually results in additional energy loss and decoherence of the system. A controlled and coherent energy redistribution process has rarely been explored for coherent addressing of magnons. The only exception to date is the Bose-Einstein condensation of magnons \cite{magnon_bec,magnon_bec1,magnon_bec2,magnon_bec3}. However, creating coherent magnon condensate requires the magnon thermalization to be faster than the magnon relaxation, thus limiting the application in low-amplitude, coherent magnon engineering.

Recently, an interesting finding of a pump-induced magnon mode suggests that the nonlinear magnon interaction can be harnessed for controlling coherent energy transduction \cite{RaoPRL23, LuWeiPRApplied23,wang2023giant}. Using a strong microwave pump, the excited magnetostatic (MS) magnon mode manifests an anticrossing feature, in which the pump power controls the level splitting. However, the microscopic origin of such a process is yet unexplored, limiting the further engineering of such a phenomenon for coherent magnon control.  

In this work, we show that this pump-induced magnon anticrossing is due to the coupling between the MS mode and the magnon condensate via a coherent three-magnon splitting and confluence process \cite{PattenPRB09,QuPRB23}. Using a comprehensive analytical model, we show that the coupling strength is proportional to the population of the magnon condensate that is generated via the nonlinear magnon scattering, in which the power threshold for observing the magnon anticrossing matches well with the power threshold of a single-tone, nonlinear magnon broadening. In addition, we reveal that a cutoff frequency exists for the pump-induced magnon anticrossing for each MS mode including the high-order ones. The cutoff frequency is closely linked to the selection rule of three-magnon scattering: above the cutoff frequency, there is no available magnon mode due to the required energy conservation, and the three-magnon scattering is therefore forbidden. Our results open up new avenues for coherent magnon interaction with nonlinnear magnonic process.

\begin{figure*}[htb]
 \centering
 \includegraphics[width=7.4 in]{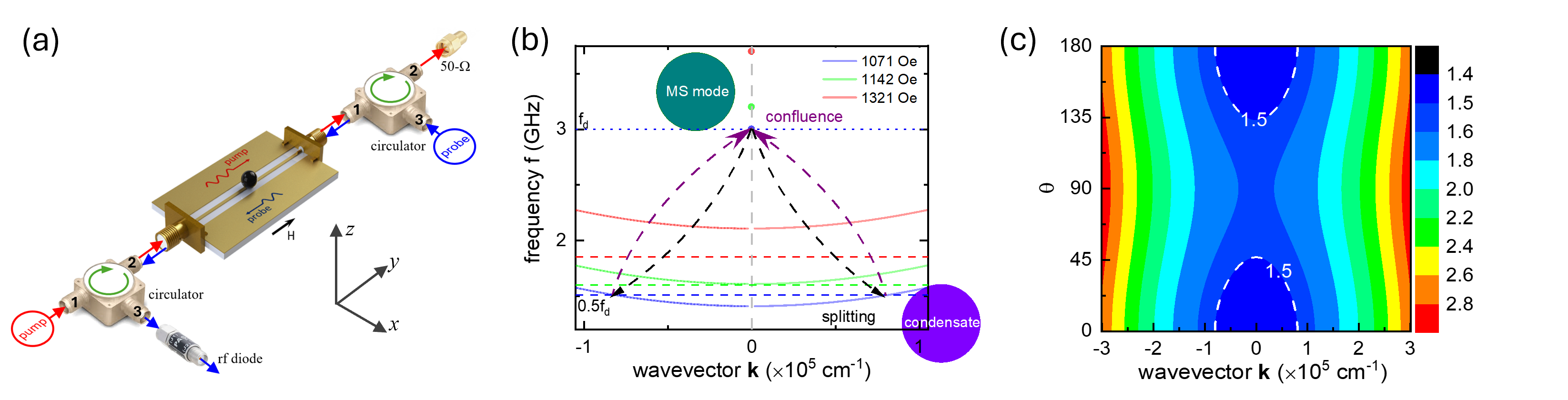} 
 \caption{(a) Schematics of the experimental setup. The counter-propagating \textcolor{black}{(enabled by a pair of broadband circulator)} pump and probe signals interact with the YIG sphere on the CPW. The spin wave excitations are detected by field-modulation FMR using a microwave diode and a lock-in amplifier. The magnetic field $H$ is applied along the signal line, satisfying the perpendicular excitation and pumping geometry. (b) Illustration of the three-magnon splitting and confluence, forming a `Rabi-like' process giving rise to coherent coupling between the condensate state and the magnetostatic mode and the avoided level repulsion. \textcolor{black}{Three representative magnon dispersions are exemplified taking the first magnetostatic mode $m$ = 1: $H$ = 1071.0 Oe (blue), 1142 Oe (green), and 1321 Oe (red), corresponding to different pump frequencies, $f_d$ = 3.0 GHz, 3.2 GHz and 3.7 GHz, respectively. Above the frequency threshold of the three-magnon scattering (red), the requirement of magnon frequency of $\frac{1}{2}f_d$ is not satisfied for the magnon wavevector $\kbf\neq0$. (c) The colormap of the intrinsic frequency $f$ of the magnon in the spherical axes of the wavevector amplitude and the polar angle $\theta$. The intrinsic frequency is degenerate in the azimuthal angle $\phi$, thus not illustrated. The intrinsic frequency at $\frac{f_d}{2}$= 1.5 GHz is plotted in the egg-shaped contour line (white dashed line) corresponding to the multiple degenerate $\kbf\neq0$ modes ($|\kbf|$,$\theta$,$\phi$),  an extension to (b) with magnetostatic frequency $f_d$=3.0 GHz in the blue line.}}  
 \label{fig1}
\end{figure*}

\section{Experiments} 

Our experimental setup is illustrated in Fig.\ref{fig1}(a): a YIG sphere with a diameter of 1.0 mm is placed on top of the signal line of a coplanar waveguide (CPW). The external magnetic field, \textit{H}, is applied in-plane (\textit{xy}) and along the signal line, corresponding to the perpendicular excitation geometry (\textit{H} $\perp$ $h_\textrm{rf}$). The system response is monitored by a weak microwave probe signal when a strong microwave pump ($f_{d}$) is applied, which induces the magnon condensate through perpendicular pumping, see Fig. \ref{fig1}(b). \textcolor{black}{This condensate is different from the Bose–Einstein condensate \cite{magnon_bec} in parametric pumping: a microwave photon creating two primary magnons and thermal relaxation of primary magnons to a quasi-equilibrium distribution of thermalized magnons in four-magnon scattering)}. The magnetostatic mode and the condensate mode, dominate our system, shown in Fig. \ref{fig1_1}(b), while the three-magnon scattering involving both splitting and confluence occurs solely between these dominant magnons, introducing the intensity of the dominant modes far deviating from their thermal level. Thus, this system is not in quasi-equilibrium and the chemical potential is not adopted to describe this Rabi-like system. Multiple MS modes are present in our experiment, as will be discussed later. 

The pump and probe signals propagate in opposite direction in the CPW by using a pair of broadband circulator at each end of the CPW. The pump path is terminated with a 50-$\Omega$ resistor to eliminate reflection. The probe signal transmitted through the device is sent to a microwave diode and detected by a lock-in amplifier using the field-modulation ferromagnetic resonance (FMR) technique \cite{inman2022hybrid}, \textcolor{black}{where a reference signal from the lock-in amplifier is amplified and fed to a modulation coil and modulates the external field with the reference frequency. The microwave diode signal is lock-in amplified at this frequency. The output of the lock-in amplifier is then proportional to the field derivative of the transmitted microwave power \cite{maksymov2015broadband}}. 

\section{Results and Discussions} 

In our system, the three-magnon splitting/confluence take place under the strong microwave pump, see Fig. \ref{fig1}(b). Each MS mode excited at $f_d$ splits into a pair of magnons in the spin wave modes at the half frequencies $f_d$/2 with opposite wavevectors, and they recombine to form a $f_d$ mode following an oscillating fashion, akin to an 'Rabi-like' oscillation in strongly coupled systems. \textcolor{black}{For a lower frequency of MS mode, the magnon dispersion satisfies $f_d$/2 at specific wavevectors. Upon increasing the frequency of the MS mode through enhancing the corresponding resonance magnetic field based on the Kittel equation, such condition breaks at a threshold frequency where only a pair of opposite wavevectors meets the frequency requirement of $f_d$/2. Above this frequency threshold, three-magnon scattering is disabled fundamentally due to no wavevector in the magnon dispersion satisfying the frequency $f_d$/2.}The Hamiltonian of such splitting/confluence process can be described as: 
\begin{equation}
    H_{3m}=\frac{1}{2} \Sigma_k(\zeta_{\kbf} c_0 c^*_{\kbf} c^*_{-\kbf} + \zeta_{\kbf}^* c_0^* c_{\kbf} c_{-\kbf} ),
\label{Eq:Hami}
\end{equation}
\noindent where $c_0$ denotes the wavevector $\kbf=0$ MS spin waves, $c_k$ the parametrically generated $\kbf\neq0$ spin waves (which eventually form the magnon condensate), and $\zeta_{\kbf}=\hbar \gamma M_s \sin(2\theta)/4$ is the three-magnon coupling coefficient, where $\hbar$ is the reduced Planck constant, $\gamma$ is the gyromagnetic ratio, $M_s$ is the magnetization saturation and $\theta$ is the angle between the wavevector $\kbf$ and the applied magnetic field, thus the maximum coupling occurs at $\theta=45^{\circ}$ when fulfilling the conditions of three-magnon scattering. Simultaneously, $\zeta_{\kbf}$ is a real number in the sphere system, where $\zeta_{\kbf}^*=\zeta_{\kbf}$. \textcolor{black}{Note that the three-magnon dispersion of the YIG sphere is slightly different from that of the YIG film \cite{ordonez2009three,liu2019time}. The colormap of the intrinsic frequency $f$ of the magnon in the spherical axes of the wavevector amplitude and the polar angle $\theta$ is illustrated in Fig. \ref{fig1}(c).} More details on this theoretical aspect is included in the Supplemental Materials \cite{sm} (see also references \cite{roschmann1977properties,fletcher1959ferrimagnetic,white1960use,walker1957magnetostatic,leo2020identification} therein).   

\begin{figure}[htb]
 \centering
 \includegraphics[width=3.4 in]{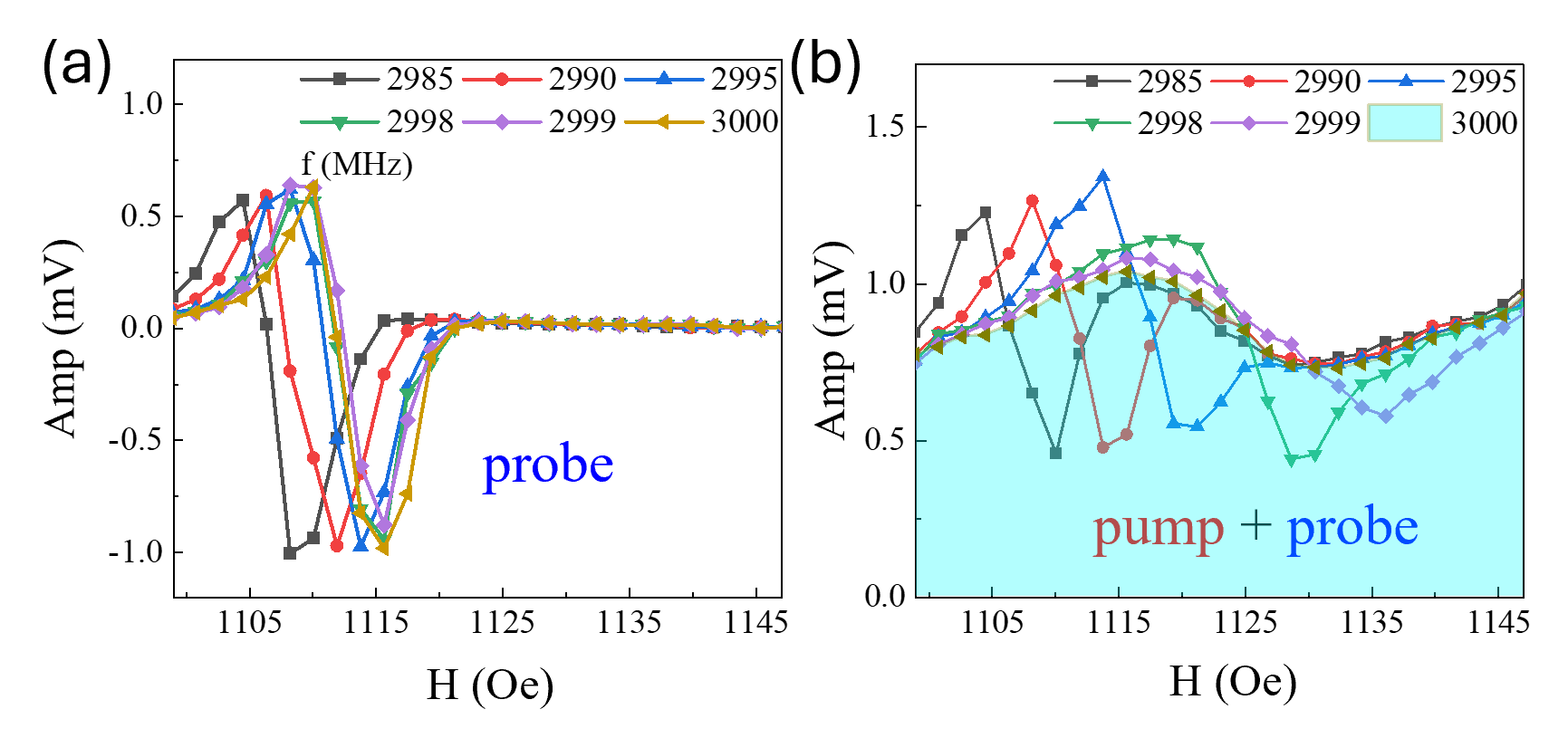}
 \caption{(a,b) The first magnetostatic mode, $m$ = 1 (centered around $H$ = 1115.0 Oe) measured at different frequencies in the situation where a pump ($f_d$ = 3.0 GHz) is off (a), or on (b). MS: magnetostatic.} 
\label{fig1_1}
\end{figure}

When the pump is weak, the magnon splitting and accordingly the intensity of the induced condensate is insignificant. However, as the pump amplitude increases, two emergent observations arise: \textit{(i)} the magnon spectra broaden significantly with much suppressed susceptibility due to the onset of the three-magnon scattering \cite{qu2020nonlinear}, and \textit{(ii)}, the confluence (accordingly, the back-action to the MS magnon mode) becomes prominent and effectively introduces an additional damping channel to the MS mode.  

\begin{figure*}[htb]
 \centering
 \includegraphics[width=6.4 in]{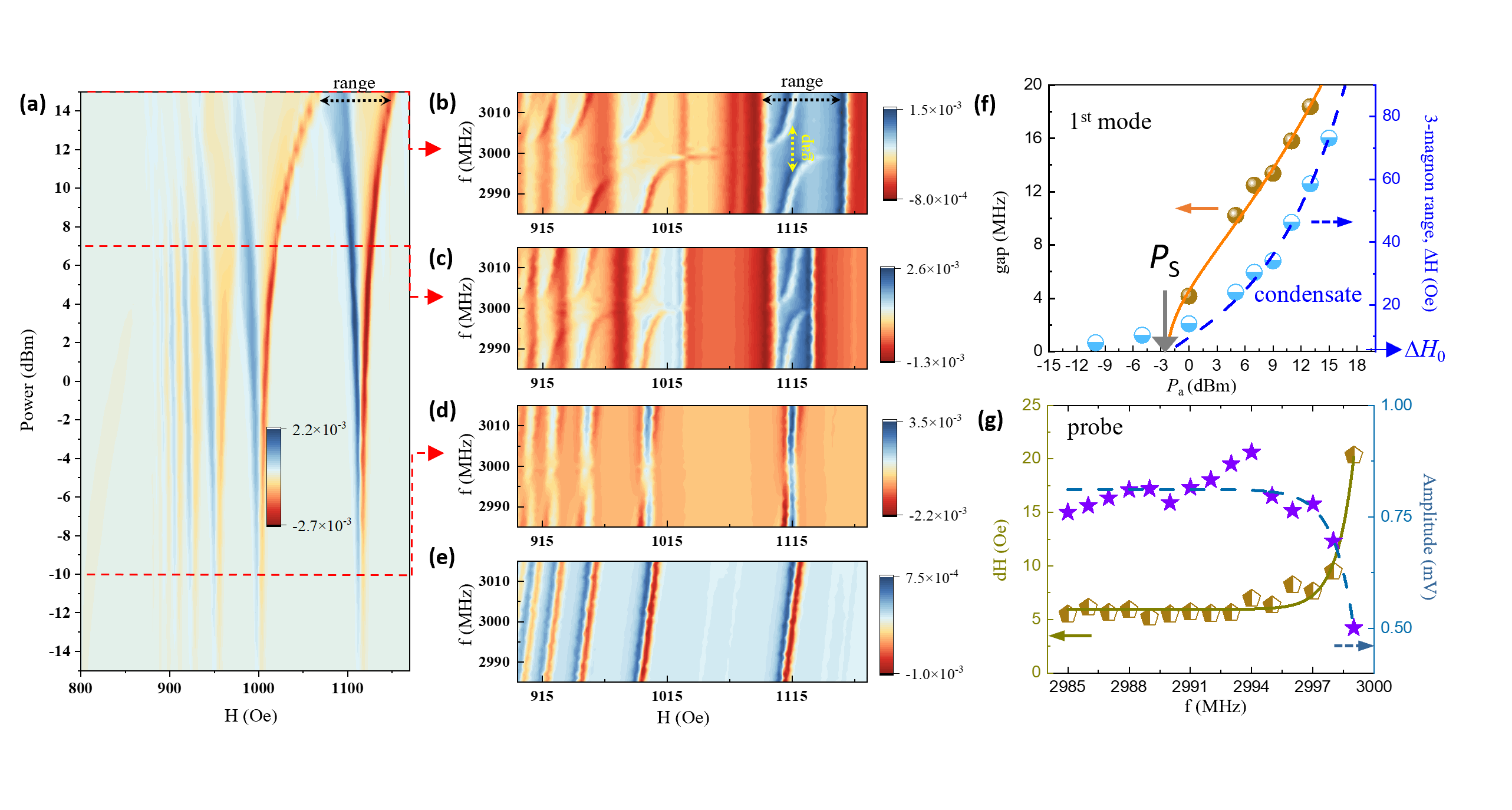}
 \caption{(a) Magnon spectra measured at 3.0 GHz \textcolor{black}{at different input microwave power and} at a broad magnetic-field range covering at least six identifiable MS modes. The $f-H$ dispersion spectra scanned near the pump frequency ($f_d = 3.0$ GHz) and at selective input power levels (b) 15 dBm, (c) 7 dBm, (d) -10 dBm, and (e) no pump. (f) The extracted anti-crossing 'gap' and condensate 'range' versus the input power for the 1$^{st}$ MS mode. A gap cannot be clearly identified for powers below 0 dBm. Solid lines are theoretical fits.  (g) The frequency-dependent linewidth and signal amplitude of the 1$^{st}$ MS mode in response to a 15 dBm pump at 3.0 GHz.  }
\label{fig2}
\end{figure*}

As an example, \textcolor{black}{Fig.\ref{fig1_1}(a,b) shows} the fundamental MS mode (centered around $H = 1115.0$ Oe) measured at different frequencies in the situation where a pump ($f_{d} = 3.0$ GHz) is either off (a) or on (b). Due to the field-modulation FMR technique and lock-in detection, the susceptibility of the whole system can be captured. When the pump is off, the signals at different frequencies correspond to the MS mode and are of similar amplitude with no amplitude offset with respect to zero. But, once the pump kicks in, the signal (at the same selected frequencies) is superimposed on a broad plateau, i.e., the continuous spin-wave band due to the onset of the three-magnon scattering. In addition, as the MS mode gets near to the pump frequency (3.0 GHz), its amplitude decreases and the lineshape broadens, caused by the additional damping channel by its nonlinear coupling to the magnon condensate.

\begin{figure}[htb]
 \centering
 \includegraphics[width=1.05\columnwidth]{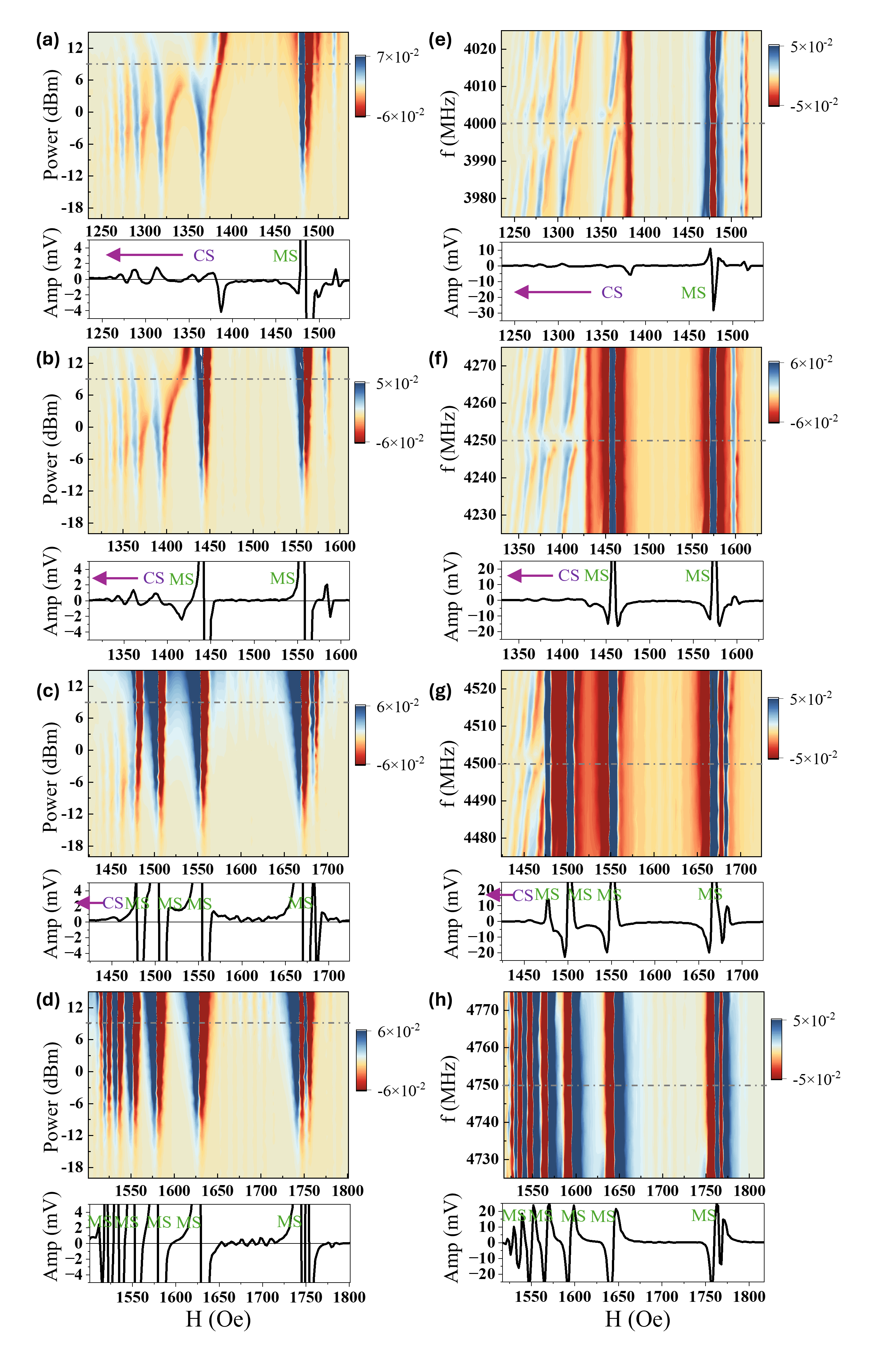}
 \caption{(a-d) Power scan contour plots at selective frequencies: (a) 4.0, (b) 4.2, (c) 4.5, (d) 4.7 GHz. A line-scan trace at a pump power of 9 dBm is shown for each contour plot. MS: magnetostatic mode. CS: condensate mode. The MS and CS modes are identified for each plot. (e-h) $f-H$ dispersion contour plots measured at a fixed pump power (9 dBm) and at selective frequencies: (e) 4.0, (f) 4.25, (g) 4.5, (h) 4.75 GHz. A line-scan trace at the pump frequency (center of anticrossing gap) is shown for each contour plot. }
 \label{fig3}
\end{figure}

The same effect takes place for multiple MS modes, which can be measured by expanding the sweeping range of the bias magnetic field, as shown in Fig. \ref{fig2}. These modes are identified as the MS modes with mode order ($m,m,0$), where $m=1,2,3,...$ is the angular mode number, and the fundamental mode ($m=1$) corresponds to the FMR mode. The evolution from a common MS mode to a condensate state by the pumping power is well exemplified by the two-dimensional (2D) contour plot in Fig. \ref{fig2}(a), measured at 3.0 GHz: at lower power level (below $-2$ dBm), individual, sharp resonances with narrow linewidths can be identified, which correspond to the genuine MS modes. However, as the power further increases, the three-magnon process becomes significant with much broadened linewidths and suppressed resonance amplitudes. From Fig.\ref{fig1}(b) it is clear that the possible wavevectors of the scattered magnons are not discrete, but instead they can take a range of values within a broad spin wave continuum. Therefore, the scattered magnons form a continuous spin wave band (the condensate), whose field span (range, $\Delta H$) can be measured and characterized directly from the 2D contour plot. 

The level of formation of such condensate has a direct impact on the MS mode by its back-action: Figures \ref{fig2}(b-e) show the $f-H$ relation contour plots measured (using a $-10$ dBm probe) at a fixed pump of $f_{d} = 3.0$ GHz and at selective power levels. When the pump is absent, Fig.\ref{fig2}(e), linear dispersion of the MS modes is observed. When a small pump is applied ($-10$ dBm), Fig.\ref{fig2}(d), the back-action of the condensate to the MS mode emerges: one can observe both a finite spin-wave band (the `range') and the MS modes, but they are almost only linearly overlapped (weakly coupled). As power further increases, Fig. \ref{fig2}(c)-(b): first, the `range' further expands; second, the back-action of the condensate to the MS mode becomes prominent, resulting in the `bending' of the MS mode spectra, akin to an anti-crossing `gap' centered around the pump ($f_d$). \textcolor{black}{Notably, such a `bending effect' only lives up to the extent confined by the condensate `range' and sharply cuts off at the boundary. This is in stark contrast with a conventional anticrossing feature observed for common hybrid photon-magnon systems, where such bounds do not exist for a conventional anti-crossing feature.}       

We develop a semi-analytical model to accurately describe the nonlinear dynamics of magnon scattering through the equation of motion, which is derived from a 1$^{st}$-order Hamiltonian including the self-energy of the magnons, and the scattering between the MS ($\kbf$=0) and spin wave ($\kbf\neq0$) magnons. The scattering process includes both forward creation process, i.e., the annihilation of one MS magnon with the creation of two spin wave magnons, and its inverse effect, the backward confluence process, i.e., the creation of one MS magnon with the annihalation of two spin wave magnons. The scattering occurs at a long timescale, e.g., sub-microsecond, compared to the short timescale of the magnetization precession in magnons, e.g., sub-nanosecond. The scattering process can be described by the dynamic equations \cite{QuPRB23}:
\begin{eqnarray}
\dot{c}_0=-\eta_0 c_0-\zeta_{\kbf} c_{\kbf}^2 + \upsilon h_a \label{eq:dc0_dt} \\
\dot{c_{\kbf}}= -\eta_k c_{\kbf}+\zeta_{\kbf} c_0 c_{k}
\label{eq:dck_dt}
\end{eqnarray}
\noindent where $c_0$ ($c_{\kbf}$) and $\eta_0$ ($\eta_k$) are the intensity and relaxation rate of the MS (spin wave) magnons, respectively. $\zeta_{\kbf}$ is the exchange strength between MS and spin wave magnon modes. $\upsilon$ is the capability of the driving force $h_a$, of the microwave field to create $\kbf=0$ magnon, where the force $h_a=\sqrt {\frac {2 P_a}{R_\mathrm{load}}}/d_\mathrm{CPW}$ is determined by the microwave power $P_a$, the load resistance $R_\mathrm{load}$ and the width of the coplanar waveguide $d_\mathrm{CPW}$.  

The constraints for these wavevectors $\kbf$ follow the laws of energy: $f_{\kbf}+f_{-\kbf}=f_{\kbf=0}$ and momentum conservation: the magnon pair of $\pm\kbf$ are created or annihilated spontaneously. The initial condition of the dynamic equations is the intensity of the thermal magnons, where their levels follow the Bose-Einstein distribution $\sim e^{-\hbar \omega_{\kbf}/k_B T}$, where $k_B$ is the Boltzmann constant and $T$ is the temperature.

The fixed point for the linear regime corresponds to:  
\begin{eqnarray}
c_{0}&=&\upsilon h_a/\eta_0 \label{eq:c0_s_linear} \\
c_{\kbf}&=&0
\label{eq:ck_s_linear}
\end{eqnarray}

In the linear regime, only $\kbf=0$ is present and the magnetization undergoes uniform precession. Thus the intensity $c_{0}$ in the linear regime is proportional to the microwave driving force $h_a$. This force is insufficient in amplitude to excite the three-magnon scattering, where $c_{\kbf}=0$ until this force enhances to the threshold $h_S=\eta_0 \eta_{\kbf}/(\zeta_{\kbf} \upsilon)$, i.e., reaching the Suhl instability. Above this threshold, the magnon scattering is triggered and a pronounced $c_{\kbf}$ is activated.  

Mathematically, the nonlinear regime’s fixed point yields the magnon intensity $c_0$ and $c_\kbf$,
\begin{eqnarray}
c_0&=&\eta_{\kbf}/\zeta_{\kbf} \label{eq:c0_s_nonlinear} \,,\\
c_{\kbf}&=&\sqrt{\frac{\upsilon h_a-\eta_0\eta_k/\zeta_\kbf}{ \zeta_\kbf}}\,,
\label{eq:ck_s_nonlinear}
\end{eqnarray}
Expressing the spin wave magnon intensity in Eq.\ref{eq:ck_s_nonlinear} within the threshold $h_S$ yields,
\begin{equation}\label{eq:ck_power}
c_{\kbf}=\sqrt{\frac{\upsilon}{\zeta_\kbf}(h_a-h_S)}\sim\sqrt{\sqrt{P_a}-\sqrt{P_S}}\,.
\end{equation}
In the nonlinear regime, the spin wave magnon presents a significant amount, much larger than its original thermal level, due to the three-magnon scattering. Thus this amount reflects the power-tuned coupling between MS and spin wave magnons, and experimentally fingerprints the gap size in Fig. \ref{fig2}(f), where the fit yields a threshold power $P_S=0.6$ mW. Correspondingly,  the three-magnon (condensate) linewidth (range, $\Delta H$), denoted as $\Delta H = 2\Delta \omega_0 /\gamma = 2 \Delta H_0 \sqrt{P_a/P_S-1}$, is shown and fitted in Fig.\ref{fig2}(f), yielding a generic magnon linewdith $\Delta H_0=5.71$ Oe.

Reciprocally, the back-action induces additional damping of the MS mode and attenuation of its amplitude, which are both confirmed in the experiment and summarized in Fig.\ref{fig2}(g), as an example for the 1$^{st}$ MS mode: the peak-to-peak linewidth ($\delta H$) increases and the signal amplitude decreases both rapidly as the MS mode frequency is closer to the pump. 

Finally, another strong evidence to our proposed mechanism comes from the concurrent frequency cutoff for both the anticrossing gap and the nonlinear broadening. Because the principles demonstrated above also apply to high order MS modes ($m>1$) which are observed in our experiment, we can examine their distinct frequency thresholds for activating such a nonlinear coupling. According to the theory of three-magnon scattering \cite{Suhl57,LiuPRB19,QuPRB23,qu2020nonlinear}, above a certain frequency threshold, the three-magnon process will be prohibited due to absence of magnon bands at the half frequency, illustrated in Fig.\ref{fig1}(b). In order to determine the threshold frequency for each MS mode, comprehensive power spectra were obtained at frequencies between 3.0 to 4.7 GHz at a step of 0.1 GHz. The threshold frequency for invoking the three-magnon process is determined as: 3.6 GHz (1$^{st}$),4.2 GHz (2$^{nd}$), 4.4 GHz (3$^{rd}$), 4.6 GHz (4$^{th}$), 4.7 GHz (5$^{th}$). Figure \ref{fig3}(a-d) show the power scan contour plot at selective frequencies, (a) 4.0, (b) 4.2, (c) 4.5, (d) 4.7 GHz. It is obvious that as the frequency increases, the magnon bands (`ranges') for each mode progressively disappear (starting from the 1$^{st}$), after passing the respective threshold frequencies. This leads to a direct voiding of the corresponding `bending effect', revealed by the $f-H$ relation measured at the cutoff frequency, in Fig.\ref{fig3}(e-h): the anti-crossing `gap' progressively disappears after reaching the cutoff frequencies. Such an observation corroborates the strong correlation between the anti-crossing `gap' and the condensate formation via three-magnon splitting/confluence. 

\section{Summary}

In summary, we report a new type of nonlinear coupling between a magnon condensate and magnetostatic modes in a YIG sphere, activated by the energy pumping from a waveguide. We show that the magnon condensate can be excited via the nonlinear three-magnon splitting processes, and its back-action to the magnetostatic mode via the confluence process is significantly enhanced by the presence of a strong pump signal, which leads to a `bending effect' of the MS mode spectra, akin to an anti-crossing `gap' in the strongly coupled magnonic systems. Our discovered new mechanism opens a new direction for developing engineerable magnonic systems leveraging coupled spin wave phenomena.  

\section{Acknowledgments}

We thank Vasyl Tyberkevych and Andrei Slavin for valuable discussions about the mechanism of three-magnon scattering. The experimental measurements about pump-induced magnon anticrossing performed at University of North Carolina at Chapel Hill are supported by the U.S. National Science Foundation  (NSF) under Grant No. ECCS-2246254. X.Z. acknowledges support from the NSF-(2337713) and the Office of Naval Research Young Investigator Program (N00014-23-1-2144). Y.L. acknowledges support from the U.S. DOE, Office of Science, Basic Energy Sciences, Materials Sciences and Engineering Division under contract No. DE-SC0022060.


\bibliography{ref}

\begin{thebibliography}{51}%
\makeatletter
\providecommand \@ifxundefined [1]{%
 \@ifx{#1\undefined}
}%
\providecommand \@ifnum [1]{%
 \ifnum #1\expandafter \@firstoftwo
 \else \expandafter \@secondoftwo
 \fi
}%
\providecommand \@ifx [1]{%
 \ifx #1\expandafter \@firstoftwo
 \else \expandafter \@secondoftwo
 \fi
}%
\providecommand \natexlab [1]{#1}%
\providecommand \enquote  [1]{``#1''}%
\providecommand \bibnamefont  [1]{#1}%
\providecommand \bibfnamefont [1]{#1}%
\providecommand \citenamefont [1]{#1}%
\providecommand \href@noop [0]{\@secondoftwo}%
\providecommand \href [0]{\begingroup \@sanitize@url \@href}%
\providecommand \@href[1]{\@@startlink{#1}\@@href}%
\providecommand \@@href[1]{\endgroup#1\@@endlink}%
\providecommand \@sanitize@url [0]{\catcode `\\12\catcode `\$12\catcode `\&12\catcode `\#12\catcode `\^12\catcode `\_12\catcode `\%12\relax}%
\providecommand \@@startlink[1]{}%
\providecommand \@@endlink[0]{}%
\providecommand \url  [0]{\begingroup\@sanitize@url \@url }%
\providecommand \@url [1]{\endgroup\@href {#1}{\urlprefix }}%
\providecommand \urlprefix  [0]{URL }%
\providecommand \Eprint [0]{\href }%
\providecommand \doibase [0]{http://dx.doi.org/}%
\providecommand \selectlanguage [0]{\@gobble}%
\providecommand \bibinfo  [0]{\@secondoftwo}%
\providecommand \bibfield  [0]{\@secondoftwo}%
\providecommand \translation [1]{[#1]}%
\providecommand \BibitemOpen [0]{}%
\providecommand \bibitemStop [0]{}%
\providecommand \bibitemNoStop [0]{.\EOS\space}%
\providecommand \EOS [0]{\spacefactor3000\relax}%
\providecommand \BibitemShut  [1]{\csname bibitem#1\endcsname}%
\let\auto@bib@innerbib\@empty
\bibitem [{\citenamefont {Lachance-Quirion}\ \emph {et~al.}(2019)\citenamefont {Lachance-Quirion}, \citenamefont {Tabuchi}, \citenamefont {Gloppe}, \citenamefont {Usami},\ and\ \citenamefont {Nakamura}}]{LachanceQuirionAPEx2019}%
  \BibitemOpen
  \bibfield  {author} {\bibinfo {author} {\bibfnamefont {D.}~\bibnamefont {Lachance-Quirion}}, \bibinfo {author} {\bibfnamefont {Y.}~\bibnamefont {Tabuchi}}, \bibinfo {author} {\bibfnamefont {A.}~\bibnamefont {Gloppe}}, \bibinfo {author} {\bibfnamefont {K.}~\bibnamefont {Usami}}, \ and\ \bibinfo {author} {\bibfnamefont {Y.}~\bibnamefont {Nakamura}},\ }\href@noop {} {\bibfield  {journal} {\bibinfo  {journal} {Appl. Phys. Express}\ }\textbf {\bibinfo {volume} {12}},\ \bibinfo {pages} {070101} (\bibinfo {year} {2019})}\BibitemShut {NoStop}%
\bibitem [{\citenamefont {Li}\ \emph {et~al.}(2020)\citenamefont {Li}, \citenamefont {Zhang}, \citenamefont {Tyberkevych}, \citenamefont {Kwok}, \citenamefont {Hoffmann},\ and\ \citenamefont {Novosad}}]{LiJAP2020}%
  \BibitemOpen
  \bibfield  {author} {\bibinfo {author} {\bibfnamefont {Y.}~\bibnamefont {Li}}, \bibinfo {author} {\bibfnamefont {W.}~\bibnamefont {Zhang}}, \bibinfo {author} {\bibfnamefont {V.}~\bibnamefont {Tyberkevych}}, \bibinfo {author} {\bibfnamefont {W.-K.}\ \bibnamefont {Kwok}}, \bibinfo {author} {\bibfnamefont {A.}~\bibnamefont {Hoffmann}}, \ and\ \bibinfo {author} {\bibfnamefont {V.}~\bibnamefont {Novosad}},\ }\href@noop {} {\bibfield  {journal} {\bibinfo  {journal} {J. Appl. Phys.}\ }\textbf {\bibinfo {volume} {128}},\ \bibinfo {pages} {130902} (\bibinfo {year} {2020})}\BibitemShut {NoStop}%
\bibitem [{\citenamefont {Awschalom}\ \emph {et~al.}(2021)\citenamefont {Awschalom}, \citenamefont {Du}, \citenamefont {He}, \citenamefont {Heremans}, \citenamefont {Hoffmann}, \citenamefont {Hou}, \citenamefont {Kurebayashi}, \citenamefont {Li}, \citenamefont {Liu}, \citenamefont {Novosad}, \citenamefont {Sklenar}, \citenamefont {Sullivan}, \citenamefont {Sun}, \citenamefont {Tang}, \citenamefont {Tyberkevych}, \citenamefont {Trevillian}, \citenamefont {Tsen}, \citenamefont {Weiss}, \citenamefont {Zhang}, \citenamefont {Zhang}, \citenamefont {Zhao},\ and\ \citenamefont {Zollitsch}}]{Awschalom21}%
  \BibitemOpen
  \bibfield  {author} {\bibinfo {author} {\bibfnamefont {D.~D.}\ \bibnamefont {Awschalom}}, \bibinfo {author} {\bibfnamefont {C.~R.}\ \bibnamefont {Du}}, \bibinfo {author} {\bibfnamefont {R.}~\bibnamefont {He}}, \bibinfo {author} {\bibfnamefont {F.~J.}\ \bibnamefont {Heremans}}, \bibinfo {author} {\bibfnamefont {A.}~\bibnamefont {Hoffmann}}, \bibinfo {author} {\bibfnamefont {J.}~\bibnamefont {Hou}}, \bibinfo {author} {\bibfnamefont {H.}~\bibnamefont {Kurebayashi}}, \bibinfo {author} {\bibfnamefont {Y.}~\bibnamefont {Li}}, \bibinfo {author} {\bibfnamefont {L.}~\bibnamefont {Liu}}, \bibinfo {author} {\bibfnamefont {V.}~\bibnamefont {Novosad}}, \bibinfo {author} {\bibfnamefont {J.}~\bibnamefont {Sklenar}}, \bibinfo {author} {\bibfnamefont {S.~E.}\ \bibnamefont {Sullivan}}, \bibinfo {author} {\bibfnamefont {D.}~\bibnamefont {Sun}}, \bibinfo {author} {\bibfnamefont {H.}~\bibnamefont {Tang}}, \bibinfo {author} {\bibfnamefont {V.}~\bibnamefont {Tyberkevych}}, \bibinfo {author} {\bibfnamefont {C.}~\bibnamefont
  {Trevillian}}, \bibinfo {author} {\bibfnamefont {A.~W.}\ \bibnamefont {Tsen}}, \bibinfo {author} {\bibfnamefont {L.~R.}\ \bibnamefont {Weiss}}, \bibinfo {author} {\bibfnamefont {W.}~\bibnamefont {Zhang}}, \bibinfo {author} {\bibfnamefont {X.}~\bibnamefont {Zhang}}, \bibinfo {author} {\bibfnamefont {L.}~\bibnamefont {Zhao}}, \ and\ \bibinfo {author} {\bibfnamefont {C.~W.}\ \bibnamefont {Zollitsch}},\ }\href@noop {} {\bibfield  {journal} {\bibinfo  {journal} {IEEE Transactions on Quantum Engineering}\ }\textbf {\bibinfo {volume} {2}},\ \bibinfo {pages} {1} (\bibinfo {year} {2021})}\BibitemShut {NoStop}%
\bibitem [{\citenamefont {Huebl}\ \emph {et~al.}(2013)\citenamefont {Huebl}, \citenamefont {Zollitsch}, \citenamefont {Lotze}, \citenamefont {Hocke}, \citenamefont {Greifenstein}, \citenamefont {Marx}, \citenamefont {Gross},\ and\ \citenamefont {Goennenwein}}]{HueblPRL2013}%
  \BibitemOpen
  \bibfield  {author} {\bibinfo {author} {\bibfnamefont {H.}~\bibnamefont {Huebl}}, \bibinfo {author} {\bibfnamefont {C.~W.}\ \bibnamefont {Zollitsch}}, \bibinfo {author} {\bibfnamefont {J.}~\bibnamefont {Lotze}}, \bibinfo {author} {\bibfnamefont {F.}~\bibnamefont {Hocke}}, \bibinfo {author} {\bibfnamefont {M.}~\bibnamefont {Greifenstein}}, \bibinfo {author} {\bibfnamefont {A.}~\bibnamefont {Marx}}, \bibinfo {author} {\bibfnamefont {R.}~\bibnamefont {Gross}}, \ and\ \bibinfo {author} {\bibfnamefont {S.~T.~B.}\ \bibnamefont {Goennenwein}},\ }\href {\doibase 10.1103/PhysRevLett.111.127003} {\bibfield  {journal} {\bibinfo  {journal} {Phys. Rev. Lett.}\ }\textbf {\bibinfo {volume} {111}},\ \bibinfo {pages} {127003} (\bibinfo {year} {2013})}\BibitemShut {NoStop}%
\bibitem [{\citenamefont {Tabuchi}\ \emph {et~al.}(2014)\citenamefont {Tabuchi}, \citenamefont {Ishino}, \citenamefont {Ishikawa}, \citenamefont {Yamazaki}, \citenamefont {Usami},\ and\ \citenamefont {Nakamura}}]{TabuchiPRL2014}%
  \BibitemOpen
  \bibfield  {author} {\bibinfo {author} {\bibfnamefont {Y.}~\bibnamefont {Tabuchi}}, \bibinfo {author} {\bibfnamefont {S.}~\bibnamefont {Ishino}}, \bibinfo {author} {\bibfnamefont {T.}~\bibnamefont {Ishikawa}}, \bibinfo {author} {\bibfnamefont {R.}~\bibnamefont {Yamazaki}}, \bibinfo {author} {\bibfnamefont {K.}~\bibnamefont {Usami}}, \ and\ \bibinfo {author} {\bibfnamefont {Y.}~\bibnamefont {Nakamura}},\ }\href {\doibase 10.1103/PhysRevLett.113.083603} {\bibfield  {journal} {\bibinfo  {journal} {Phys. Rev. Lett.}\ }\textbf {\bibinfo {volume} {113}},\ \bibinfo {pages} {083603} (\bibinfo {year} {2014})}\BibitemShut {NoStop}%
\bibitem [{\citenamefont {Zhang}\ \emph {et~al.}(2014)\citenamefont {Zhang}, \citenamefont {Zou}, \citenamefont {Jiang},\ and\ \citenamefont {Tang}}]{ZhangPRL2014}%
  \BibitemOpen
  \bibfield  {author} {\bibinfo {author} {\bibfnamefont {X.}~\bibnamefont {Zhang}}, \bibinfo {author} {\bibfnamefont {C.-L.}\ \bibnamefont {Zou}}, \bibinfo {author} {\bibfnamefont {L.}~\bibnamefont {Jiang}}, \ and\ \bibinfo {author} {\bibfnamefont {H.~X.}\ \bibnamefont {Tang}},\ }\href {\doibase 10.1103/PhysRevLett.113.156401} {\bibfield  {journal} {\bibinfo  {journal} {Phys. Rev. Lett.}\ }\textbf {\bibinfo {volume} {113}},\ \bibinfo {pages} {156401} (\bibinfo {year} {2014})}\BibitemShut {NoStop}%
\bibitem [{\citenamefont {Goryachev}\ \emph {et~al.}(2014)\citenamefont {Goryachev}, \citenamefont {Farr}, \citenamefont {Creedon}, \citenamefont {Fan}, \citenamefont {Kostylev},\ and\ \citenamefont {Tobar}}]{GoryachevPRApplied2014}%
  \BibitemOpen
  \bibfield  {author} {\bibinfo {author} {\bibfnamefont {M.}~\bibnamefont {Goryachev}}, \bibinfo {author} {\bibfnamefont {W.~G.}\ \bibnamefont {Farr}}, \bibinfo {author} {\bibfnamefont {D.~L.}\ \bibnamefont {Creedon}}, \bibinfo {author} {\bibfnamefont {Y.}~\bibnamefont {Fan}}, \bibinfo {author} {\bibfnamefont {M.}~\bibnamefont {Kostylev}}, \ and\ \bibinfo {author} {\bibfnamefont {M.~E.}\ \bibnamefont {Tobar}},\ }\href {\doibase 10.1103/PhysRevApplied.2.054002} {\bibfield  {journal} {\bibinfo  {journal} {Phys. Rev. Applied}\ }\textbf {\bibinfo {volume} {2}},\ \bibinfo {pages} {054002} (\bibinfo {year} {2014})}\BibitemShut {NoStop}%
\bibitem [{\citenamefont {Bai}\ \emph {et~al.}(2015)\citenamefont {Bai}, \citenamefont {Harder}, \citenamefont {Chen}, \citenamefont {Fan}, \citenamefont {Xiao},\ and\ \citenamefont {Hu}}]{BaiPRL2015}%
  \BibitemOpen
  \bibfield  {author} {\bibinfo {author} {\bibfnamefont {L.}~\bibnamefont {Bai}}, \bibinfo {author} {\bibfnamefont {M.}~\bibnamefont {Harder}}, \bibinfo {author} {\bibfnamefont {Y.~P.}\ \bibnamefont {Chen}}, \bibinfo {author} {\bibfnamefont {X.}~\bibnamefont {Fan}}, \bibinfo {author} {\bibfnamefont {J.~Q.}\ \bibnamefont {Xiao}}, \ and\ \bibinfo {author} {\bibfnamefont {C.-M.}\ \bibnamefont {Hu}},\ }\href {\doibase 10.1103/PhysRevLett.114.227201} {\bibfield  {journal} {\bibinfo  {journal} {Phys. Rev. Lett.}\ }\textbf {\bibinfo {volume} {114}},\ \bibinfo {pages} {227201} (\bibinfo {year} {2015})}\BibitemShut {NoStop}%
\bibitem [{\citenamefont {Li}\ \emph {et~al.}(2019)\citenamefont {Li}, \citenamefont {Polakovic}, \citenamefont {Wang}, \citenamefont {Xu}, \citenamefont {Lendinez}, \citenamefont {Zhang}, \citenamefont {Ding}, \citenamefont {Khaire}, \citenamefont {Saglam}, \citenamefont {Divan}, \citenamefont {Pearson}, \citenamefont {Kwok}, \citenamefont {Xiao}, \citenamefont {Novosad}, \citenamefont {Hoffmann},\ and\ \citenamefont {Zhang}}]{LiPRL2019_magnon}%
  \BibitemOpen
  \bibfield  {author} {\bibinfo {author} {\bibfnamefont {Y.}~\bibnamefont {Li}}, \bibinfo {author} {\bibfnamefont {T.}~\bibnamefont {Polakovic}}, \bibinfo {author} {\bibfnamefont {Y.-L.}\ \bibnamefont {Wang}}, \bibinfo {author} {\bibfnamefont {J.}~\bibnamefont {Xu}}, \bibinfo {author} {\bibfnamefont {S.}~\bibnamefont {Lendinez}}, \bibinfo {author} {\bibfnamefont {Z.}~\bibnamefont {Zhang}}, \bibinfo {author} {\bibfnamefont {J.}~\bibnamefont {Ding}}, \bibinfo {author} {\bibfnamefont {T.}~\bibnamefont {Khaire}}, \bibinfo {author} {\bibfnamefont {H.}~\bibnamefont {Saglam}}, \bibinfo {author} {\bibfnamefont {R.}~\bibnamefont {Divan}}, \bibinfo {author} {\bibfnamefont {J.}~\bibnamefont {Pearson}}, \bibinfo {author} {\bibfnamefont {W.-K.}\ \bibnamefont {Kwok}}, \bibinfo {author} {\bibfnamefont {Z.}~\bibnamefont {Xiao}}, \bibinfo {author} {\bibfnamefont {V.}~\bibnamefont {Novosad}}, \bibinfo {author} {\bibfnamefont {A.}~\bibnamefont {Hoffmann}}, \ and\ \bibinfo {author} {\bibfnamefont {W.}~\bibnamefont {Zhang}},\
  }\href {\doibase 10.1103/PhysRevLett.123.107701} {\bibfield  {journal} {\bibinfo  {journal} {Phys. Rev. Lett.}\ }\textbf {\bibinfo {volume} {123}},\ \bibinfo {pages} {107701} (\bibinfo {year} {2019})}\BibitemShut {NoStop}%
\bibitem [{\citenamefont {Hou}\ and\ \citenamefont {Liu}(2019)}]{HouPRL2019}%
  \BibitemOpen
  \bibfield  {author} {\bibinfo {author} {\bibfnamefont {J.~T.}\ \bibnamefont {Hou}}\ and\ \bibinfo {author} {\bibfnamefont {L.}~\bibnamefont {Liu}},\ }\href {\doibase 10.1103/PhysRevLett.123.107702} {\bibfield  {journal} {\bibinfo  {journal} {Phys. Rev. Lett.}\ }\textbf {\bibinfo {volume} {123}},\ \bibinfo {pages} {107702} (\bibinfo {year} {2019})}\BibitemShut {NoStop}%
\bibitem [{\citenamefont {Harder}\ \emph {et~al.}(2018)\citenamefont {Harder}, \citenamefont {Yang}, \citenamefont {Yao}, \citenamefont {Yu}, \citenamefont {Rao}, \citenamefont {Gui}, \citenamefont {Stamps},\ and\ \citenamefont {Hu}}]{HarderPRL2018}%
  \BibitemOpen
  \bibfield  {author} {\bibinfo {author} {\bibfnamefont {M.}~\bibnamefont {Harder}}, \bibinfo {author} {\bibfnamefont {Y.}~\bibnamefont {Yang}}, \bibinfo {author} {\bibfnamefont {B.~M.}\ \bibnamefont {Yao}}, \bibinfo {author} {\bibfnamefont {C.~H.}\ \bibnamefont {Yu}}, \bibinfo {author} {\bibfnamefont {J.~W.}\ \bibnamefont {Rao}}, \bibinfo {author} {\bibfnamefont {Y.~S.}\ \bibnamefont {Gui}}, \bibinfo {author} {\bibfnamefont {R.~L.}\ \bibnamefont {Stamps}}, \ and\ \bibinfo {author} {\bibfnamefont {C.-M.}\ \bibnamefont {Hu}},\ }\href {\doibase 10.1103/PhysRevLett.121.137203} {\bibfield  {journal} {\bibinfo  {journal} {Phys. Rev. Lett.}\ }\textbf {\bibinfo {volume} {121}},\ \bibinfo {pages} {137203} (\bibinfo {year} {2018})}\BibitemShut {NoStop}%
\bibitem [{\citenamefont {Wang}\ \emph {et~al.}(2019)\citenamefont {Wang}, \citenamefont {Rao}, \citenamefont {Yang}, \citenamefont {Xu}, \citenamefont {Gui}, \citenamefont {Yao}, \citenamefont {You},\ and\ \citenamefont {Hu}}]{WangPRL19}%
  \BibitemOpen
  \bibfield  {author} {\bibinfo {author} {\bibfnamefont {Y.-P.}\ \bibnamefont {Wang}}, \bibinfo {author} {\bibfnamefont {J.~W.}\ \bibnamefont {Rao}}, \bibinfo {author} {\bibfnamefont {Y.}~\bibnamefont {Yang}}, \bibinfo {author} {\bibfnamefont {P.-C.}\ \bibnamefont {Xu}}, \bibinfo {author} {\bibfnamefont {Y.~S.}\ \bibnamefont {Gui}}, \bibinfo {author} {\bibfnamefont {B.~M.}\ \bibnamefont {Yao}}, \bibinfo {author} {\bibfnamefont {J.~Q.}\ \bibnamefont {You}}, \ and\ \bibinfo {author} {\bibfnamefont {C.-M.}\ \bibnamefont {Hu}},\ }\href {\doibase 10.1103/PhysRevLett.123.127202} {\bibfield  {journal} {\bibinfo  {journal} {Phys. Rev. Lett.}\ }\textbf {\bibinfo {volume} {123}},\ \bibinfo {pages} {127202} (\bibinfo {year} {2019})}\BibitemShut {NoStop}%
\bibitem [{\citenamefont {Zhang}\ \emph {et~al.}(2020)\citenamefont {Zhang}, \citenamefont {Galda}, \citenamefont {Han}, \citenamefont {Jin},\ and\ \citenamefont {Vinokur}}]{ZhangPRApplied19}%
  \BibitemOpen
  \bibfield  {author} {\bibinfo {author} {\bibfnamefont {X.}~\bibnamefont {Zhang}}, \bibinfo {author} {\bibfnamefont {A.}~\bibnamefont {Galda}}, \bibinfo {author} {\bibfnamefont {X.}~\bibnamefont {Han}}, \bibinfo {author} {\bibfnamefont {D.}~\bibnamefont {Jin}}, \ and\ \bibinfo {author} {\bibfnamefont {V.~M.}\ \bibnamefont {Vinokur}},\ }\href {\doibase 10.1103/PhysRevApplied.13.044039} {\bibfield  {journal} {\bibinfo  {journal} {Phys. Rev. Appl.}\ }\textbf {\bibinfo {volume} {13}},\ \bibinfo {pages} {044039} (\bibinfo {year} {2020})}\BibitemShut {NoStop}%
\bibitem [{\citenamefont {Zhang}\ \emph {et~al.}(2017)\citenamefont {Zhang}, \citenamefont {Luo}, \citenamefont {Wang}, \citenamefont {Li},\ and\ \citenamefont {You}}]{YouNComm17}%
  \BibitemOpen
  \bibfield  {author} {\bibinfo {author} {\bibfnamefont {D.}~\bibnamefont {Zhang}}, \bibinfo {author} {\bibfnamefont {X.-Q.}\ \bibnamefont {Luo}}, \bibinfo {author} {\bibfnamefont {Y.-P.}\ \bibnamefont {Wang}}, \bibinfo {author} {\bibfnamefont {T.-F.}\ \bibnamefont {Li}}, \ and\ \bibinfo {author} {\bibfnamefont {J.~Q.}\ \bibnamefont {You}},\ }\href@noop {} {\bibfield  {journal} {\bibinfo  {journal} {Nature Comm.}\ }\textbf {\bibinfo {volume} {8}},\ \bibinfo {pages} {1368} (\bibinfo {year} {2017})}\BibitemShut {NoStop}%
\bibitem [{\citenamefont {Zhang}\ \emph {et~al.}(2019)\citenamefont {Zhang}, \citenamefont {Ding}, \citenamefont {Zhou}, \citenamefont {Xu},\ and\ \citenamefont {Jin}}]{ZhangPRL19}%
  \BibitemOpen
  \bibfield  {author} {\bibinfo {author} {\bibfnamefont {X.}~\bibnamefont {Zhang}}, \bibinfo {author} {\bibfnamefont {K.}~\bibnamefont {Ding}}, \bibinfo {author} {\bibfnamefont {X.}~\bibnamefont {Zhou}}, \bibinfo {author} {\bibfnamefont {J.}~\bibnamefont {Xu}}, \ and\ \bibinfo {author} {\bibfnamefont {D.}~\bibnamefont {Jin}},\ }\href {\doibase 10.1103/PhysRevLett.123.237202} {\bibfield  {journal} {\bibinfo  {journal} {Phys. Rev. Lett.}\ }\textbf {\bibinfo {volume} {123}},\ \bibinfo {pages} {237202} (\bibinfo {year} {2019})}\BibitemShut {NoStop}%
\bibitem [{\citenamefont {Xu}\ \emph {et~al.}(2020)\citenamefont {Xu}, \citenamefont {Zhong}, \citenamefont {Han}, \citenamefont {Jin}, \citenamefont {Jiang},\ and\ \citenamefont {Zhang}}]{ZhangPRL20}%
  \BibitemOpen
  \bibfield  {author} {\bibinfo {author} {\bibfnamefont {J.}~\bibnamefont {Xu}}, \bibinfo {author} {\bibfnamefont {C.}~\bibnamefont {Zhong}}, \bibinfo {author} {\bibfnamefont {X.}~\bibnamefont {Han}}, \bibinfo {author} {\bibfnamefont {D.}~\bibnamefont {Jin}}, \bibinfo {author} {\bibfnamefont {L.}~\bibnamefont {Jiang}}, \ and\ \bibinfo {author} {\bibfnamefont {X.}~\bibnamefont {Zhang}},\ }\href {\doibase 10.1103/PhysRevLett.125.237201} {\bibfield  {journal} {\bibinfo  {journal} {Phys. Rev. Lett.}\ }\textbf {\bibinfo {volume} {125}},\ \bibinfo {pages} {237201} (\bibinfo {year} {2020})}\BibitemShut {NoStop}%
\bibitem [{\citenamefont {Xiong}\ \emph {et~al.}(2024)\citenamefont {Xiong}, \citenamefont {Christy}, \citenamefont {Dong}, \citenamefont {Comstock}, \citenamefont {Sun}, \citenamefont {Li}, \citenamefont {Cahoon}, \citenamefont {Yang},\ and\ \citenamefont {Zhang}}]{xiong2024combinatorial}%
  \BibitemOpen
  \bibfield  {author} {\bibinfo {author} {\bibfnamefont {Y.}~\bibnamefont {Xiong}}, \bibinfo {author} {\bibfnamefont {A.}~\bibnamefont {Christy}}, \bibinfo {author} {\bibfnamefont {Y.}~\bibnamefont {Dong}}, \bibinfo {author} {\bibfnamefont {A.~H.}\ \bibnamefont {Comstock}}, \bibinfo {author} {\bibfnamefont {D.}~\bibnamefont {Sun}}, \bibinfo {author} {\bibfnamefont {Y.}~\bibnamefont {Li}}, \bibinfo {author} {\bibfnamefont {J.~F.}\ \bibnamefont {Cahoon}}, \bibinfo {author} {\bibfnamefont {B.}~\bibnamefont {Yang}}, \ and\ \bibinfo {author} {\bibfnamefont {W.}~\bibnamefont {Zhang}},\ }\href@noop {} {\bibfield  {journal} {\bibinfo  {journal} {Physical Review Applied}\ }\textbf {\bibinfo {volume} {21}},\ \bibinfo {pages} {034034} (\bibinfo {year} {2024})}\BibitemShut {NoStop}%
\bibitem [{\citenamefont {Lachance-Quirion}\ \emph {et~al.}(2020)\citenamefont {Lachance-Quirion}, \citenamefont {Wolski}, \citenamefont {Tabuchi}, \citenamefont {Kono}, \citenamefont {Usami},\ and\ \citenamefont {Nakamura}}]{LachanceQuirionScience2020}%
  \BibitemOpen
  \bibfield  {author} {\bibinfo {author} {\bibfnamefont {D.}~\bibnamefont {Lachance-Quirion}}, \bibinfo {author} {\bibfnamefont {S.~P.}\ \bibnamefont {Wolski}}, \bibinfo {author} {\bibfnamefont {Y.}~\bibnamefont {Tabuchi}}, \bibinfo {author} {\bibfnamefont {S.}~\bibnamefont {Kono}}, \bibinfo {author} {\bibfnamefont {K.}~\bibnamefont {Usami}}, \ and\ \bibinfo {author} {\bibfnamefont {Y.}~\bibnamefont {Nakamura}},\ }\href@noop {} {\bibfield  {journal} {\bibinfo  {journal} {Science}\ }\textbf {\bibinfo {volume} {367}},\ \bibinfo {pages} {425} (\bibinfo {year} {2020})}\BibitemShut {NoStop}%
\bibitem [{\citenamefont {Xu}\ \emph {et~al.}(2023)\citenamefont {Xu}, \citenamefont {Gu}, \citenamefont {Li}, \citenamefont {Weng}, \citenamefont {Wang}, \citenamefont {Li}, \citenamefont {Wang}, \citenamefont {Zhu},\ and\ \citenamefont {You}}]{YouPRL23}%
  \BibitemOpen
  \bibfield  {author} {\bibinfo {author} {\bibfnamefont {D.}~\bibnamefont {Xu}}, \bibinfo {author} {\bibfnamefont {X.-K.}\ \bibnamefont {Gu}}, \bibinfo {author} {\bibfnamefont {H.-K.}\ \bibnamefont {Li}}, \bibinfo {author} {\bibfnamefont {Y.-C.}\ \bibnamefont {Weng}}, \bibinfo {author} {\bibfnamefont {Y.-P.}\ \bibnamefont {Wang}}, \bibinfo {author} {\bibfnamefont {J.}~\bibnamefont {Li}}, \bibinfo {author} {\bibfnamefont {H.}~\bibnamefont {Wang}}, \bibinfo {author} {\bibfnamefont {S.-Y.}\ \bibnamefont {Zhu}}, \ and\ \bibinfo {author} {\bibfnamefont {J.~Q.}\ \bibnamefont {You}},\ }\href {\doibase 10.1103/PhysRevLett.130.193603} {\bibfield  {journal} {\bibinfo  {journal} {Phys. Rev. Lett.}\ }\textbf {\bibinfo {volume} {130}},\ \bibinfo {pages} {193603} (\bibinfo {year} {2023})}\BibitemShut {NoStop}%
\bibitem [{\citenamefont {Yuan}\ \emph {et~al.}(2022)\citenamefont {Yuan}, \citenamefont {Cao}, \citenamefont {Kamra}, \citenamefont {Duine},\ and\ \citenamefont {Yan}}]{YuanPhysRep22}%
  \BibitemOpen
  \bibfield  {author} {\bibinfo {author} {\bibfnamefont {H.}~\bibnamefont {Yuan}}, \bibinfo {author} {\bibfnamefont {Y.}~\bibnamefont {Cao}}, \bibinfo {author} {\bibfnamefont {A.}~\bibnamefont {Kamra}}, \bibinfo {author} {\bibfnamefont {R.~A.}\ \bibnamefont {Duine}}, \ and\ \bibinfo {author} {\bibfnamefont {P.}~\bibnamefont {Yan}},\ }\href@noop {} {\bibfield  {journal} {\bibinfo  {journal} {Physics Reports}\ }\textbf {\bibinfo {volume} {965}},\ \bibinfo {pages} {1} (\bibinfo {year} {2022})}\BibitemShut {NoStop}%
\bibitem [{\citenamefont {Flebus}\ \emph {et~al.}(2024)\citenamefont {Flebus}, \citenamefont {Grundler}, \citenamefont {Rana}, \citenamefont {Otani}, \citenamefont {Barsukov}, \citenamefont {Barman}, \citenamefont {Gubbiotti}, \citenamefont {Landeros}, \citenamefont {Akerman}, \citenamefont {Ebels} \emph {et~al.}}]{flebus20242024}%
  \BibitemOpen
  \bibfield  {author} {\bibinfo {author} {\bibfnamefont {B.}~\bibnamefont {Flebus}}, \bibinfo {author} {\bibfnamefont {D.}~\bibnamefont {Grundler}}, \bibinfo {author} {\bibfnamefont {B.}~\bibnamefont {Rana}}, \bibinfo {author} {\bibfnamefont {Y.}~\bibnamefont {Otani}}, \bibinfo {author} {\bibfnamefont {I.}~\bibnamefont {Barsukov}}, \bibinfo {author} {\bibfnamefont {A.}~\bibnamefont {Barman}}, \bibinfo {author} {\bibfnamefont {G.}~\bibnamefont {Gubbiotti}}, \bibinfo {author} {\bibfnamefont {P.}~\bibnamefont {Landeros}}, \bibinfo {author} {\bibfnamefont {J.}~\bibnamefont {Akerman}}, \bibinfo {author} {\bibfnamefont {U.~S.}\ \bibnamefont {Ebels}},  \emph {et~al.},\ }\href@noop {} {\bibfield  {journal} {\bibinfo  {journal} {Journal of Physics: Condensed Matter}\ } (\bibinfo {year} {2024})}\BibitemShut {NoStop}%
\bibitem [{\citenamefont {Zheng}\ \emph {et~al.}(2023)\citenamefont {Zheng}, \citenamefont {Wang}, \citenamefont {Wang}, \citenamefont {Sun}, \citenamefont {He}, \citenamefont {Yan},\ and\ \citenamefont {Yuan}}]{YuanJAP23}%
  \BibitemOpen
  \bibfield  {author} {\bibinfo {author} {\bibfnamefont {S.}~\bibnamefont {Zheng}}, \bibinfo {author} {\bibfnamefont {Z.}~\bibnamefont {Wang}}, \bibinfo {author} {\bibfnamefont {Y.}~\bibnamefont {Wang}}, \bibinfo {author} {\bibfnamefont {F.}~\bibnamefont {Sun}}, \bibinfo {author} {\bibfnamefont {Q.}~\bibnamefont {He}}, \bibinfo {author} {\bibfnamefont {P.}~\bibnamefont {Yan}}, \ and\ \bibinfo {author} {\bibfnamefont {H.~Y.}\ \bibnamefont {Yuan}},\ }\href@noop {} {\bibfield  {journal} {\bibinfo  {journal} {J. Appl. Phys.}\ }\textbf {\bibinfo {volume} {134}},\ \bibinfo {pages} {151101} (\bibinfo {year} {2023})}\BibitemShut {NoStop}%
\bibitem [{\citenamefont {Makiuchi}\ \emph {et~al.}(2024)\citenamefont {Makiuchi}, \citenamefont {Hioki}, \citenamefont {Shimizu}, \citenamefont {Hoshi}, \citenamefont {Elyasi}, \citenamefont {Yamamoto}, \citenamefont {Yokoi}, \citenamefont {Serga}, \citenamefont {Hillebrands}, \citenamefont {Bauer} \emph {et~al.}}]{makiuchi2024persistent}%
  \BibitemOpen
  \bibfield  {author} {\bibinfo {author} {\bibfnamefont {T.}~\bibnamefont {Makiuchi}}, \bibinfo {author} {\bibfnamefont {T.}~\bibnamefont {Hioki}}, \bibinfo {author} {\bibfnamefont {H.}~\bibnamefont {Shimizu}}, \bibinfo {author} {\bibfnamefont {K.}~\bibnamefont {Hoshi}}, \bibinfo {author} {\bibfnamefont {M.}~\bibnamefont {Elyasi}}, \bibinfo {author} {\bibfnamefont {K.}~\bibnamefont {Yamamoto}}, \bibinfo {author} {\bibfnamefont {N.}~\bibnamefont {Yokoi}}, \bibinfo {author} {\bibfnamefont {A.}~\bibnamefont {Serga}}, \bibinfo {author} {\bibfnamefont {B.}~\bibnamefont {Hillebrands}}, \bibinfo {author} {\bibfnamefont {G.}~\bibnamefont {Bauer}},  \emph {et~al.},\ }\href@noop {} {\bibfield  {journal} {\bibinfo  {journal} {Nature Materials}\ ,\ \bibinfo {pages} {1}} (\bibinfo {year} {2024})}\BibitemShut {NoStop}%
\bibitem [{\citenamefont {Mathieu}\ \emph {et~al.}(2003)\citenamefont {Mathieu}, \citenamefont {Synogatch},\ and\ \citenamefont {Patton}}]{PattonPRB03}%
  \BibitemOpen
  \bibfield  {author} {\bibinfo {author} {\bibfnamefont {C.}~\bibnamefont {Mathieu}}, \bibinfo {author} {\bibfnamefont {V.~T.}\ \bibnamefont {Synogatch}}, \ and\ \bibinfo {author} {\bibfnamefont {C.~E.}\ \bibnamefont {Patton}},\ }\href {\doibase 10.1103/PhysRevB.67.104402} {\bibfield  {journal} {\bibinfo  {journal} {Phys. Rev. B}\ }\textbf {\bibinfo {volume} {67}},\ \bibinfo {pages} {104402} (\bibinfo {year} {2003})}\BibitemShut {NoStop}%
\bibitem [{\citenamefont {Wojewoda}\ \emph {et~al.}(2023)\citenamefont {Wojewoda}, \citenamefont {Ligmajer}, \citenamefont {Hrto{\v{n}}}, \citenamefont {Kl{\'\i}ma}, \citenamefont {Dhankhar}, \citenamefont {Dav{\'\i}dkov{\'a}}, \citenamefont {Sta{\v{n}}o}, \citenamefont {Holobr{\'a}dek}, \citenamefont {Kr{\v{c}}ma}, \citenamefont {Zl{\'a}mal} \emph {et~al.}}]{wojewoda2023observing}%
  \BibitemOpen
  \bibfield  {author} {\bibinfo {author} {\bibfnamefont {O.}~\bibnamefont {Wojewoda}}, \bibinfo {author} {\bibfnamefont {F.}~\bibnamefont {Ligmajer}}, \bibinfo {author} {\bibfnamefont {M.}~\bibnamefont {Hrto{\v{n}}}}, \bibinfo {author} {\bibfnamefont {J.}~\bibnamefont {Kl{\'\i}ma}}, \bibinfo {author} {\bibfnamefont {M.}~\bibnamefont {Dhankhar}}, \bibinfo {author} {\bibfnamefont {K.}~\bibnamefont {Dav{\'\i}dkov{\'a}}}, \bibinfo {author} {\bibfnamefont {M.}~\bibnamefont {Sta{\v{n}}o}}, \bibinfo {author} {\bibfnamefont {J.}~\bibnamefont {Holobr{\'a}dek}}, \bibinfo {author} {\bibfnamefont {J.}~\bibnamefont {Kr{\v{c}}ma}}, \bibinfo {author} {\bibfnamefont {J.}~\bibnamefont {Zl{\'a}mal}},  \emph {et~al.},\ }\href@noop {} {\bibfield  {journal} {\bibinfo  {journal} {Comms. Phys.}\ }\textbf {\bibinfo {volume} {6}},\ \bibinfo {pages} {94} (\bibinfo {year} {2023})}\BibitemShut {NoStop}%
\bibitem [{\citenamefont {K\"orber}\ \emph {et~al.}(2020)\citenamefont {K\"orber}, \citenamefont {Schultheiss}, \citenamefont {Hula}, \citenamefont {Verba}, \citenamefont {Fassbender}, \citenamefont {K\'akay},\ and\ \citenamefont {Schultheiss}}]{PRL207203}%
  \BibitemOpen
  \bibfield  {author} {\bibinfo {author} {\bibfnamefont {L.}~\bibnamefont {K\"orber}}, \bibinfo {author} {\bibfnamefont {K.}~\bibnamefont {Schultheiss}}, \bibinfo {author} {\bibfnamefont {T.}~\bibnamefont {Hula}}, \bibinfo {author} {\bibfnamefont {R.}~\bibnamefont {Verba}}, \bibinfo {author} {\bibfnamefont {J.}~\bibnamefont {Fassbender}}, \bibinfo {author} {\bibfnamefont {A.}~\bibnamefont {K\'akay}}, \ and\ \bibinfo {author} {\bibfnamefont {H.}~\bibnamefont {Schultheiss}},\ }\href {\doibase 10.1103/PhysRevLett.125.207203} {\bibfield  {journal} {\bibinfo  {journal} {Phys. Rev. Lett.}\ }\textbf {\bibinfo {volume} {125}},\ \bibinfo {pages} {207203} (\bibinfo {year} {2020})}\BibitemShut {NoStop}%
\bibitem [{\citenamefont {K{\"o}rber}\ \emph {et~al.}(2023)\citenamefont {K{\"o}rber}, \citenamefont {Heins}, \citenamefont {Hula}, \citenamefont {Kim}, \citenamefont {Thlang}, \citenamefont {Schultheiss}, \citenamefont {Fassbender},\ and\ \citenamefont {Schultheiss}}]{korber2023pattern}%
  \BibitemOpen
  \bibfield  {author} {\bibinfo {author} {\bibfnamefont {L.}~\bibnamefont {K{\"o}rber}}, \bibinfo {author} {\bibfnamefont {C.}~\bibnamefont {Heins}}, \bibinfo {author} {\bibfnamefont {T.}~\bibnamefont {Hula}}, \bibinfo {author} {\bibfnamefont {J.-V.}\ \bibnamefont {Kim}}, \bibinfo {author} {\bibfnamefont {S.}~\bibnamefont {Thlang}}, \bibinfo {author} {\bibfnamefont {H.}~\bibnamefont {Schultheiss}}, \bibinfo {author} {\bibfnamefont {J.}~\bibnamefont {Fassbender}}, \ and\ \bibinfo {author} {\bibfnamefont {K.}~\bibnamefont {Schultheiss}},\ }\href@noop {} {\bibfield  {journal} {\bibinfo  {journal} {Nat. Commun.}\ }\textbf {\bibinfo {volume} {14}},\ \bibinfo {pages} {3954} (\bibinfo {year} {2023})}\BibitemShut {NoStop}%
\bibitem [{\citenamefont {Ord\'o\~nez Romero}\ \emph {et~al.}(2009)\citenamefont {Ord\'o\~nez Romero}, \citenamefont {Kalinikos}, \citenamefont {Krivosik}, \citenamefont {Tong}, \citenamefont {Kabos},\ and\ \citenamefont {Patton}}]{PattenPRB09}%
  \BibitemOpen
  \bibfield  {author} {\bibinfo {author} {\bibfnamefont {C.~L.}\ \bibnamefont {Ord\'o\~nez Romero}}, \bibinfo {author} {\bibfnamefont {B.~A.}\ \bibnamefont {Kalinikos}}, \bibinfo {author} {\bibfnamefont {P.}~\bibnamefont {Krivosik}}, \bibinfo {author} {\bibfnamefont {W.}~\bibnamefont {Tong}}, \bibinfo {author} {\bibfnamefont {P.}~\bibnamefont {Kabos}}, \ and\ \bibinfo {author} {\bibfnamefont {C.~E.}\ \bibnamefont {Patton}},\ }\href {\doibase 10.1103/PhysRevB.79.144428} {\bibfield  {journal} {\bibinfo  {journal} {Phys. Rev. B}\ }\textbf {\bibinfo {volume} {79}},\ \bibinfo {pages} {144428} (\bibinfo {year} {2009})}\BibitemShut {NoStop}%
\bibitem [{\citenamefont {Kurebayashi}\ \emph {et~al.}(2011)\citenamefont {Kurebayashi}, \citenamefont {Dzyapko}, \citenamefont {Demidov}, \citenamefont {Fang}, \citenamefont {Ferguson},\ and\ \citenamefont {Demokritov}}]{Kurebayashi11}%
  \BibitemOpen
  \bibfield  {author} {\bibinfo {author} {\bibfnamefont {H.}~\bibnamefont {Kurebayashi}}, \bibinfo {author} {\bibfnamefont {O.}~\bibnamefont {Dzyapko}}, \bibinfo {author} {\bibfnamefont {V.~E.}\ \bibnamefont {Demidov}}, \bibinfo {author} {\bibfnamefont {D.}~\bibnamefont {Fang}}, \bibinfo {author} {\bibfnamefont {A.~J.}\ \bibnamefont {Ferguson}}, \ and\ \bibinfo {author} {\bibfnamefont {S.~O.}\ \bibnamefont {Demokritov}},\ }\href@noop {} {\bibfield  {journal} {\bibinfo  {journal} {Nature Mater.}\ }\textbf {\bibinfo {volume} {10}},\ \bibinfo {pages} {660} (\bibinfo {year} {2011})}\BibitemShut {NoStop}%
\bibitem [{\citenamefont {Liu}\ \emph {et~al.}(2019)\citenamefont {Liu}, \citenamefont {Riley}, \citenamefont {Ord\'o\~nez Romero}, \citenamefont {Kalinikos},\ and\ \citenamefont {Buchanan}}]{LiuPRB19}%
  \BibitemOpen
  \bibfield  {author} {\bibinfo {author} {\bibfnamefont {H.~J.~J.}\ \bibnamefont {Liu}}, \bibinfo {author} {\bibfnamefont {G.~A.}\ \bibnamefont {Riley}}, \bibinfo {author} {\bibfnamefont {C.~L.}\ \bibnamefont {Ord\'o\~nez Romero}}, \bibinfo {author} {\bibfnamefont {B.~A.}\ \bibnamefont {Kalinikos}}, \ and\ \bibinfo {author} {\bibfnamefont {K.~S.}\ \bibnamefont {Buchanan}},\ }\href {\doibase 10.1103/PhysRevB.99.024429} {\bibfield  {journal} {\bibinfo  {journal} {Phys. Rev. B}\ }\textbf {\bibinfo {volume} {99}},\ \bibinfo {pages} {024429} (\bibinfo {year} {2019})}\BibitemShut {NoStop}%
\bibitem [{\citenamefont {Schultheiss}\ \emph {et~al.}(2009)\citenamefont {Schultheiss}, \citenamefont {Janssens}, \citenamefont {van Kampen}, \citenamefont {Ciubotaru}, \citenamefont {Hermsdoerfer}, \citenamefont {Obry}, \citenamefont {Laraoui}, \citenamefont {Serga}, \citenamefont {Lagae}, \citenamefont {Slavin}, \citenamefont {Leven},\ and\ \citenamefont {Hillebrands}}]{SchultheissPRL09}%
  \BibitemOpen
  \bibfield  {author} {\bibinfo {author} {\bibfnamefont {H.}~\bibnamefont {Schultheiss}}, \bibinfo {author} {\bibfnamefont {X.}~\bibnamefont {Janssens}}, \bibinfo {author} {\bibfnamefont {M.}~\bibnamefont {van Kampen}}, \bibinfo {author} {\bibfnamefont {F.}~\bibnamefont {Ciubotaru}}, \bibinfo {author} {\bibfnamefont {S.~J.}\ \bibnamefont {Hermsdoerfer}}, \bibinfo {author} {\bibfnamefont {B.}~\bibnamefont {Obry}}, \bibinfo {author} {\bibfnamefont {A.}~\bibnamefont {Laraoui}}, \bibinfo {author} {\bibfnamefont {A.~A.}\ \bibnamefont {Serga}}, \bibinfo {author} {\bibfnamefont {L.}~\bibnamefont {Lagae}}, \bibinfo {author} {\bibfnamefont {A.~N.}\ \bibnamefont {Slavin}}, \bibinfo {author} {\bibfnamefont {B.}~\bibnamefont {Leven}}, \ and\ \bibinfo {author} {\bibfnamefont {B.}~\bibnamefont {Hillebrands}},\ }\href {\doibase 10.1103/PhysRevLett.103.157202} {\bibfield  {journal} {\bibinfo  {journal} {Phys. Rev. Lett.}\ }\textbf {\bibinfo {volume} {103}},\ \bibinfo {pages} {157202} (\bibinfo {year} {2009})}\BibitemShut
  {NoStop}%
\bibitem [{\citenamefont {Suhl}(1957)}]{Suhl57}%
  \BibitemOpen
  \bibfield  {author} {\bibinfo {author} {\bibfnamefont {H.}~\bibnamefont {Suhl}},\ }\href@noop {} {\bibfield  {journal} {\bibinfo  {journal} {J. Phys. Chem. Solids}\ }\textbf {\bibinfo {volume} {1}},\ \bibinfo {pages} {209} (\bibinfo {year} {1957})}\BibitemShut {NoStop}%
\bibitem [{\citenamefont {Demokritov}\ \emph {et~al.}(2006)\citenamefont {Demokritov}, \citenamefont {Demidov}, \citenamefont {Dzyapko}, \citenamefont {Melkov}, \citenamefont {Serga}, \citenamefont {Hillebrands},\ and\ \citenamefont {Slavin}}]{magnon_bec}%
  \BibitemOpen
  \bibfield  {author} {\bibinfo {author} {\bibfnamefont {S.~O.}\ \bibnamefont {Demokritov}}, \bibinfo {author} {\bibfnamefont {V.~E.}\ \bibnamefont {Demidov}}, \bibinfo {author} {\bibfnamefont {O.}~\bibnamefont {Dzyapko}}, \bibinfo {author} {\bibfnamefont {G.~A.}\ \bibnamefont {Melkov}}, \bibinfo {author} {\bibfnamefont {A.~A.}\ \bibnamefont {Serga}}, \bibinfo {author} {\bibfnamefont {B.}~\bibnamefont {Hillebrands}}, \ and\ \bibinfo {author} {\bibfnamefont {A.~N.}\ \bibnamefont {Slavin}},\ }\href@noop {} {\bibfield  {journal} {\bibinfo  {journal} {Nature}\ }\textbf {\bibinfo {volume} {443}},\ \bibinfo {pages} {430} (\bibinfo {year} {2006})}\BibitemShut {NoStop}%
\bibitem [{\citenamefont {Kreil}\ \emph {et~al.}(2018)\citenamefont {Kreil}, \citenamefont {Bozhko}, \citenamefont {Musiienko-Shmarova}, \citenamefont {Vasyuchka}, \citenamefont {L'vov}, \citenamefont {Pomyalov}, \citenamefont {Hillebrands},\ and\ \citenamefont {Serga}}]{magnon_bec1}%
  \BibitemOpen
  \bibfield  {author} {\bibinfo {author} {\bibfnamefont {A.~J.~E.}\ \bibnamefont {Kreil}}, \bibinfo {author} {\bibfnamefont {D.~A.}\ \bibnamefont {Bozhko}}, \bibinfo {author} {\bibfnamefont {H.~Y.}\ \bibnamefont {Musiienko-Shmarova}}, \bibinfo {author} {\bibfnamefont {V.~I.}\ \bibnamefont {Vasyuchka}}, \bibinfo {author} {\bibfnamefont {V.~S.}\ \bibnamefont {L'vov}}, \bibinfo {author} {\bibfnamefont {A.}~\bibnamefont {Pomyalov}}, \bibinfo {author} {\bibfnamefont {B.}~\bibnamefont {Hillebrands}}, \ and\ \bibinfo {author} {\bibfnamefont {A.~A.}\ \bibnamefont {Serga}},\ }\href {\doibase 10.1103/PhysRevLett.121.077203} {\bibfield  {journal} {\bibinfo  {journal} {Phys. Rev. Lett.}\ }\textbf {\bibinfo {volume} {121}},\ \bibinfo {pages} {077203} (\bibinfo {year} {2018})}\BibitemShut {NoStop}%
\bibitem [{\citenamefont {Schneider}\ \emph {et~al.}(2021)\citenamefont {Schneider}, \citenamefont {Breitbach}, \citenamefont {Serha}, \citenamefont {Wang}, \citenamefont {Serga}, \citenamefont {Slavin}, \citenamefont {Tiberkevich}, \citenamefont {Heinz}, \citenamefont {L\"agel}, \citenamefont {Br\"acher}, \citenamefont {Dubs}, \citenamefont {Knauer}, \citenamefont {Dobrovolskiy}, \citenamefont {Pirro}, \citenamefont {Hillebrands},\ and\ \citenamefont {Chumak}}]{magnon_bec2}%
  \BibitemOpen
  \bibfield  {author} {\bibinfo {author} {\bibfnamefont {M.}~\bibnamefont {Schneider}}, \bibinfo {author} {\bibfnamefont {D.}~\bibnamefont {Breitbach}}, \bibinfo {author} {\bibfnamefont {R.~O.}\ \bibnamefont {Serha}}, \bibinfo {author} {\bibfnamefont {Q.}~\bibnamefont {Wang}}, \bibinfo {author} {\bibfnamefont {A.~A.}\ \bibnamefont {Serga}}, \bibinfo {author} {\bibfnamefont {A.~N.}\ \bibnamefont {Slavin}}, \bibinfo {author} {\bibfnamefont {V.~S.}\ \bibnamefont {Tiberkevich}}, \bibinfo {author} {\bibfnamefont {B.}~\bibnamefont {Heinz}}, \bibinfo {author} {\bibfnamefont {B.}~\bibnamefont {L\"agel}}, \bibinfo {author} {\bibfnamefont {T.}~\bibnamefont {Br\"acher}}, \bibinfo {author} {\bibfnamefont {C.}~\bibnamefont {Dubs}}, \bibinfo {author} {\bibfnamefont {S.}~\bibnamefont {Knauer}}, \bibinfo {author} {\bibfnamefont {O.~V.}\ \bibnamefont {Dobrovolskiy}}, \bibinfo {author} {\bibfnamefont {P.}~\bibnamefont {Pirro}}, \bibinfo {author} {\bibfnamefont {B.}~\bibnamefont {Hillebrands}}, \ and\ \bibinfo {author}
  {\bibfnamefont {A.~V.}\ \bibnamefont {Chumak}},\ }\href {\doibase 10.1103/PhysRevLett.127.237203} {\bibfield  {journal} {\bibinfo  {journal} {Phys. Rev. Lett.}\ }\textbf {\bibinfo {volume} {127}},\ \bibinfo {pages} {237203} (\bibinfo {year} {2021})}\BibitemShut {NoStop}%
\bibitem [{\citenamefont {Mohseni}\ \emph {et~al.}(2020)\citenamefont {Mohseni}, \citenamefont {Qaiumzadeh}, \citenamefont {Serga}, \citenamefont {Brataas}, \citenamefont {Hillebrands},\ and\ \citenamefont {Pirro}}]{magnon_bec3}%
  \BibitemOpen
  \bibfield  {author} {\bibinfo {author} {\bibfnamefont {M.}~\bibnamefont {Mohseni}}, \bibinfo {author} {\bibfnamefont {A.}~\bibnamefont {Qaiumzadeh}}, \bibinfo {author} {\bibfnamefont {A.~A.}\ \bibnamefont {Serga}}, \bibinfo {author} {\bibfnamefont {A.}~\bibnamefont {Brataas}}, \bibinfo {author} {\bibfnamefont {B.}~\bibnamefont {Hillebrands}}, \ and\ \bibinfo {author} {\bibfnamefont {P.}~\bibnamefont {Pirro}},\ }\href@noop {} {\bibfield  {journal} {\bibinfo  {journal} {New J. Phys.}\ }\textbf {\bibinfo {volume} {22}},\ \bibinfo {pages} {083080} (\bibinfo {year} {2020})}\BibitemShut {NoStop}%
\bibitem [{\citenamefont {Rao}\ \emph {et~al.}(2023)\citenamefont {Rao}, \citenamefont {Yao}, \citenamefont {Wang}, \citenamefont {Zhang}, \citenamefont {Yu},\ and\ \citenamefont {Lu}}]{RaoPRL23}%
  \BibitemOpen
  \bibfield  {author} {\bibinfo {author} {\bibfnamefont {J.~W.}\ \bibnamefont {Rao}}, \bibinfo {author} {\bibfnamefont {B.}~\bibnamefont {Yao}}, \bibinfo {author} {\bibfnamefont {C.~Y.}\ \bibnamefont {Wang}}, \bibinfo {author} {\bibfnamefont {C.}~\bibnamefont {Zhang}}, \bibinfo {author} {\bibfnamefont {T.}~\bibnamefont {Yu}}, \ and\ \bibinfo {author} {\bibfnamefont {W.}~\bibnamefont {Lu}},\ }\href {\doibase 10.1103/PhysRevLett.130.046705} {\bibfield  {journal} {\bibinfo  {journal} {Phys. Rev. Lett.}\ }\textbf {\bibinfo {volume} {130}},\ \bibinfo {pages} {046705} (\bibinfo {year} {2023})}\BibitemShut {NoStop}%
\bibitem [{\citenamefont {Zhang}\ \emph {et~al.}(2023)\citenamefont {Zhang}, \citenamefont {Rao}, \citenamefont {Wang}, \citenamefont {Chen}, \citenamefont {Zhao}, \citenamefont {Yao}, \citenamefont {Xu},\ and\ \citenamefont {Lu}}]{LuWeiPRApplied23}%
  \BibitemOpen
  \bibfield  {author} {\bibinfo {author} {\bibfnamefont {C.}~\bibnamefont {Zhang}}, \bibinfo {author} {\bibfnamefont {J.}~\bibnamefont {Rao}}, \bibinfo {author} {\bibfnamefont {C.}~\bibnamefont {Wang}}, \bibinfo {author} {\bibfnamefont {Z.}~\bibnamefont {Chen}}, \bibinfo {author} {\bibfnamefont {K.}~\bibnamefont {Zhao}}, \bibinfo {author} {\bibfnamefont {B.}~\bibnamefont {Yao}}, \bibinfo {author} {\bibfnamefont {X.-G.}\ \bibnamefont {Xu}}, \ and\ \bibinfo {author} {\bibfnamefont {W.}~\bibnamefont {Lu}},\ }\href {\doibase 10.1103/PhysRevApplied.20.024074} {\bibfield  {journal} {\bibinfo  {journal} {Phys. Rev. Appl.}\ }\textbf {\bibinfo {volume} {20}},\ \bibinfo {pages} {024074} (\bibinfo {year} {2023})}\BibitemShut {NoStop}%
\bibitem [{\citenamefont {Wang}\ \emph {et~al.}(2024)\citenamefont {Wang}, \citenamefont {Rao}, \citenamefont {Chen}, \citenamefont {Zhao}, \citenamefont {Sun}, \citenamefont {Yao}, \citenamefont {Yu}, \citenamefont {Wang},\ and\ \citenamefont {Lu}}]{wang2023giant}%
  \BibitemOpen
  \bibfield  {author} {\bibinfo {author} {\bibfnamefont {C.}~\bibnamefont {Wang}}, \bibinfo {author} {\bibfnamefont {J.}~\bibnamefont {Rao}}, \bibinfo {author} {\bibfnamefont {Z.}~\bibnamefont {Chen}}, \bibinfo {author} {\bibfnamefont {K.}~\bibnamefont {Zhao}}, \bibinfo {author} {\bibfnamefont {L.}~\bibnamefont {Sun}}, \bibinfo {author} {\bibfnamefont {B.}~\bibnamefont {Yao}}, \bibinfo {author} {\bibfnamefont {T.}~\bibnamefont {Yu}}, \bibinfo {author} {\bibfnamefont {Y.-P.}\ \bibnamefont {Wang}}, \ and\ \bibinfo {author} {\bibfnamefont {W.}~\bibnamefont {Lu}},\ }\href@noop {} {\bibfield  {journal} {\bibinfo  {journal} {Nature Physics}\ }\textbf {\bibinfo {volume} {20}},\ \bibinfo {pages} {1139} (\bibinfo {year} {2024})}\BibitemShut {NoStop}%
\bibitem [{\citenamefont {Qu}\ \emph {et~al.}(2023)\citenamefont {Qu}, \citenamefont {Hamill}, \citenamefont {Victora},\ and\ \citenamefont {Crowell}}]{QuPRB23}%
  \BibitemOpen
  \bibfield  {author} {\bibinfo {author} {\bibfnamefont {T.}~\bibnamefont {Qu}}, \bibinfo {author} {\bibfnamefont {A.}~\bibnamefont {Hamill}}, \bibinfo {author} {\bibfnamefont {R.~H.}\ \bibnamefont {Victora}}, \ and\ \bibinfo {author} {\bibfnamefont {P.~A.}\ \bibnamefont {Crowell}},\ }\href {\doibase 10.1103/PhysRevB.107.L060401} {\bibfield  {journal} {\bibinfo  {journal} {Phys. Rev. B}\ }\textbf {\bibinfo {volume} {107}},\ \bibinfo {pages} {L060401} (\bibinfo {year} {2023})}\BibitemShut {NoStop}%
\bibitem [{\citenamefont {Inman}\ \emph {et~al.}(2022)\citenamefont {Inman}, \citenamefont {Xiong}, \citenamefont {Bidthanapally}, \citenamefont {Louis}, \citenamefont {Tyberkevych}, \citenamefont {Qu}, \citenamefont {Sklenar}, \citenamefont {Novosad}, \citenamefont {Li}, \citenamefont {Zhang} \emph {et~al.}}]{inman2022hybrid}%
  \BibitemOpen
  \bibfield  {author} {\bibinfo {author} {\bibfnamefont {J.}~\bibnamefont {Inman}}, \bibinfo {author} {\bibfnamefont {Y.}~\bibnamefont {Xiong}}, \bibinfo {author} {\bibfnamefont {R.}~\bibnamefont {Bidthanapally}}, \bibinfo {author} {\bibfnamefont {S.}~\bibnamefont {Louis}}, \bibinfo {author} {\bibfnamefont {V.}~\bibnamefont {Tyberkevych}}, \bibinfo {author} {\bibfnamefont {H.}~\bibnamefont {Qu}}, \bibinfo {author} {\bibfnamefont {J.}~\bibnamefont {Sklenar}}, \bibinfo {author} {\bibfnamefont {V.}~\bibnamefont {Novosad}}, \bibinfo {author} {\bibfnamefont {Y.}~\bibnamefont {Li}}, \bibinfo {author} {\bibfnamefont {X.}~\bibnamefont {Zhang}},  \emph {et~al.},\ }\href@noop {} {\bibfield  {journal} {\bibinfo  {journal} {Physical Review Applied}\ }\textbf {\bibinfo {volume} {17}},\ \bibinfo {pages} {044034} (\bibinfo {year} {2022})}\BibitemShut {NoStop}%
\bibitem [{\citenamefont {Maksymov}\ and\ \citenamefont {Kostylev}(2015)}]{maksymov2015broadband}%
  \BibitemOpen
  \bibfield  {author} {\bibinfo {author} {\bibfnamefont {I.~S.}\ \bibnamefont {Maksymov}}\ and\ \bibinfo {author} {\bibfnamefont {M.}~\bibnamefont {Kostylev}},\ }\href@noop {} {\bibfield  {journal} {\bibinfo  {journal} {Physica E: Low-dimensional Systems and Nanostructures}\ }\textbf {\bibinfo {volume} {69}},\ \bibinfo {pages} {253} (\bibinfo {year} {2015})}\BibitemShut {NoStop}%
\bibitem [{\citenamefont {\textcolor{black}{Ord{\'o}{\~n}ez-Romero, C{\'e}sar L and Kalinikos, Boris A and Krivosik, Pavol and Tong, Wei and Kabos, Pavel and Patton, Carl E}}(2009)}]{ordonez2009three}%
  \BibitemOpen
  \bibfield  {author} {\bibinfo {author} {\bibnamefont {\textcolor{black}{Ord{\'o}{\~n}ez-Romero, C{\'e}sar L and Kalinikos, Boris A and Krivosik, Pavol and Tong, Wei and Kabos, Pavel and Patton, Carl E}}},\ }\href@noop {} {\bibfield  {journal} {\bibinfo  {journal} {Physical Review B—Condensed Matter and Materials Physics}\ }\textbf {\bibinfo {volume} {79}},\ \bibinfo {pages} {144428} (\bibinfo {year} {2009})}\BibitemShut {NoStop}%
\bibitem [{\citenamefont {\textcolor{black}{Liu, HJ Jason and Riley, Grant A and Ord{\'o}{\~n}ez-Romero, C{\'e}sar L and Kalinikos, Boris A and Buchanan, Kristen S}}(2019)}]{liu2019time}%
  \BibitemOpen
  \bibfield  {author} {\bibinfo {author} {\bibnamefont {\textcolor{black}{Liu, HJ Jason and Riley, Grant A and Ord{\'o}{\~n}ez-Romero, C{\'e}sar L and Kalinikos, Boris A and Buchanan, Kristen S}}},\ }\href@noop {} {\bibfield  {journal} {\bibinfo  {journal} {Physical Review B}\ }\textbf {\bibinfo {volume} {99}},\ \bibinfo {pages} {024429} (\bibinfo {year} {2019})}\BibitemShut {NoStop}%
\bibitem [{sm()}]{sm}%
  \BibitemOpen
  \href@noop {} {}\bibinfo {note} {See Supplemental Material at [URL will be inserted by publisher] for the detailed theoretical calculations of magnon dispersion and the nonlinear coupling strength.}\BibitemShut {Stop}%
\bibitem [{\citenamefont {R{\"o}schmann}\ and\ \citenamefont {D{\"o}tsch}(1977)}]{roschmann1977properties}%
  \BibitemOpen
  \bibfield  {author} {\bibinfo {author} {\bibfnamefont {P.}~\bibnamefont {R{\"o}schmann}}\ and\ \bibinfo {author} {\bibfnamefont {H.}~\bibnamefont {D{\"o}tsch}},\ }\href@noop {} {\bibfield  {journal} {\bibinfo  {journal} {physica status solidi (b)}\ }\textbf {\bibinfo {volume} {82}},\ \bibinfo {pages} {11} (\bibinfo {year} {1977})}\BibitemShut {NoStop}%
\bibitem [{\citenamefont {Fletcher}\ and\ \citenamefont {Bell}(1959)}]{fletcher1959ferrimagnetic}%
  \BibitemOpen
  \bibfield  {author} {\bibinfo {author} {\bibfnamefont {P.}~\bibnamefont {Fletcher}}\ and\ \bibinfo {author} {\bibfnamefont {R.}~\bibnamefont {Bell}},\ }\href@noop {} {\bibfield  {journal} {\bibinfo  {journal} {Journal of applied physics}\ }\textbf {\bibinfo {volume} {30}},\ \bibinfo {pages} {687} (\bibinfo {year} {1959})}\BibitemShut {NoStop}%
\bibitem [{\citenamefont {White}(1960)}]{white1960use}%
  \BibitemOpen
  \bibfield  {author} {\bibinfo {author} {\bibfnamefont {R.~L.}\ \bibnamefont {White}},\ }\href@noop {} {\bibfield  {journal} {\bibinfo  {journal} {Journal of Applied Physics}\ }\textbf {\bibinfo {volume} {31}},\ \bibinfo {pages} {S86} (\bibinfo {year} {1960})}\BibitemShut {NoStop}%
\bibitem [{\citenamefont {Walker}(1957)}]{walker1957magnetostatic}%
  \BibitemOpen
  \bibfield  {author} {\bibinfo {author} {\bibfnamefont {L.~R.}\ \bibnamefont {Walker}},\ }\href@noop {} {\bibfield  {journal} {\bibinfo  {journal} {Physical Review}\ }\textbf {\bibinfo {volume} {105}},\ \bibinfo {pages} {390} (\bibinfo {year} {1957})}\BibitemShut {NoStop}%
\bibitem [{\citenamefont {Leo}\ \emph {et~al.}(2020)\citenamefont {Leo}, \citenamefont {Monteduro}, \citenamefont {Rizzato}, \citenamefont {Martina},\ and\ \citenamefont {Maruccio}}]{leo2020identification}%
  \BibitemOpen
  \bibfield  {author} {\bibinfo {author} {\bibfnamefont {A.}~\bibnamefont {Leo}}, \bibinfo {author} {\bibfnamefont {A.~G.}\ \bibnamefont {Monteduro}}, \bibinfo {author} {\bibfnamefont {S.}~\bibnamefont {Rizzato}}, \bibinfo {author} {\bibfnamefont {L.}~\bibnamefont {Martina}}, \ and\ \bibinfo {author} {\bibfnamefont {G.}~\bibnamefont {Maruccio}},\ }\href@noop {} {\bibfield  {journal} {\bibinfo  {journal} {Physical Review B}\ }\textbf {\bibinfo {volume} {101}},\ \bibinfo {pages} {014439} (\bibinfo {year} {2020})}\BibitemShut {NoStop}%
\bibitem [{\citenamefont {Qu}\ \emph {et~al.}(2020)\citenamefont {Qu}, \citenamefont {Venugopal}, \citenamefont {Etheridge}, \citenamefont {Peria}, \citenamefont {Srinivasan}, \citenamefont {Stadler}, \citenamefont {Crowell},\ and\ \citenamefont {Victora}}]{qu2020nonlinear}%
  \BibitemOpen
  \bibfield  {author} {\bibinfo {author} {\bibfnamefont {T.}~\bibnamefont {Qu}}, \bibinfo {author} {\bibfnamefont {A.}~\bibnamefont {Venugopal}}, \bibinfo {author} {\bibfnamefont {J.~M.}\ \bibnamefont {Etheridge}}, \bibinfo {author} {\bibfnamefont {W.~K.}\ \bibnamefont {Peria}}, \bibinfo {author} {\bibfnamefont {K.}~\bibnamefont {Srinivasan}}, \bibinfo {author} {\bibfnamefont {B.~J.}\ \bibnamefont {Stadler}}, \bibinfo {author} {\bibfnamefont {P.~A.}\ \bibnamefont {Crowell}}, \ and\ \bibinfo {author} {\bibfnamefont {R.}~\bibnamefont {Victora}},\ }\href@noop {} {\bibfield  {journal} {\bibinfo  {journal} {IEEE Access}\ }\textbf {\bibinfo {volume} {8}},\ \bibinfo {pages} {216960} (\bibinfo {year} {2020})}\BibitemShut {NoStop}%
\end{thebibliography}%


\newpage

\onecolumngrid

\newpage

\LARGE{\textbf{Supplemental Materials:}} \newline

\large{for \textbf{Pump-induced magnon anticrossing due to three-magnon splitting and confluence}} \newline
\newline
\normalsize
\textit{by} Tao Qu, Yuzan Xiong, Xufeng Zhang, and Yi Li, Wei Zhang. 
\newline

\renewcommand{\theequation}{S-\arabic{equation}}
\setcounter{equation}{0}  
\renewcommand{\thefigure}{S-\arabic{figure}}
\setcounter{figure}{0}  


\section{\\A: Experimental setup}

\begin{figure}[htb]
 \centering
 \includegraphics[width=6 in]{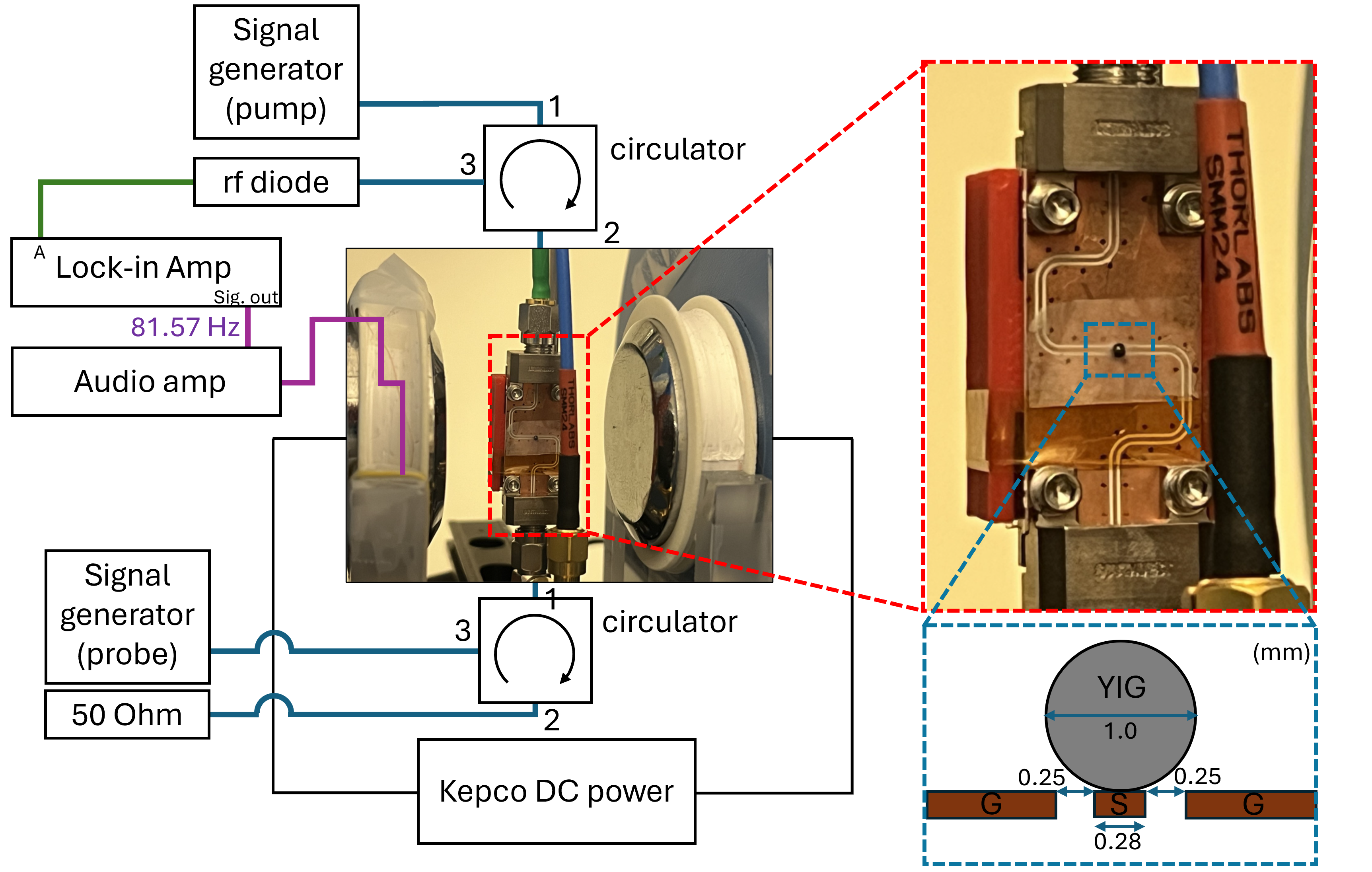}
 \caption{\textcolor{black}{Detailed experimental setup and the dimensions of the CPW and YIG sphere. }}
 \label{fig:setup_scheme}
\end{figure}

\textcolor{black}{A detailed experimental setup and the dimensions of the CPW are illustrated in Fig.\ref{fig:setup_scheme}. We used a field-modulation FMR technique, in which a slow modulation frequency (81.57 Hz) is sourced by a lock-in amplifier (sig. out) and further amplified by an audio amplifier, before it is sent to the modulation coil. The external magnetic field, \textit{H}, is applied along the signal line, corresponding to the perpendicular excitation geometry (\textit{H} $\perp$ $h_\textrm{rf}$). The system response is monitored by a weak microwave signal (probe) when another strong microwave signal (pump) is applied, using respective microwave signal generators combined with a pair of microwave circulators. The pump and probe signals propagate in opposite direction in the CPW. The pump path is terminated with a 50-$\Omega$ resistor to eliminate reflection. The probe signal transmitted through the device is sent to a microwave diode and detected by the lock-in amplifier at the same modulated coil frequency.} 

\textcolor{black}{The CPW has a signal-line width of 0.28 mm, and gap size of 0.25 mm. The dielectric layer is 0.356 mm, and the permittivity is 11. The YIG has a nominal diameter of 1.0 mm. The CPW effectively generates a non-uniform rf magnetic field to the YIG sphere, assisting the excitations of the higher-order Walker modes.}

\begin{figure}[htb]
 \centering
 \includegraphics[width=6 in]{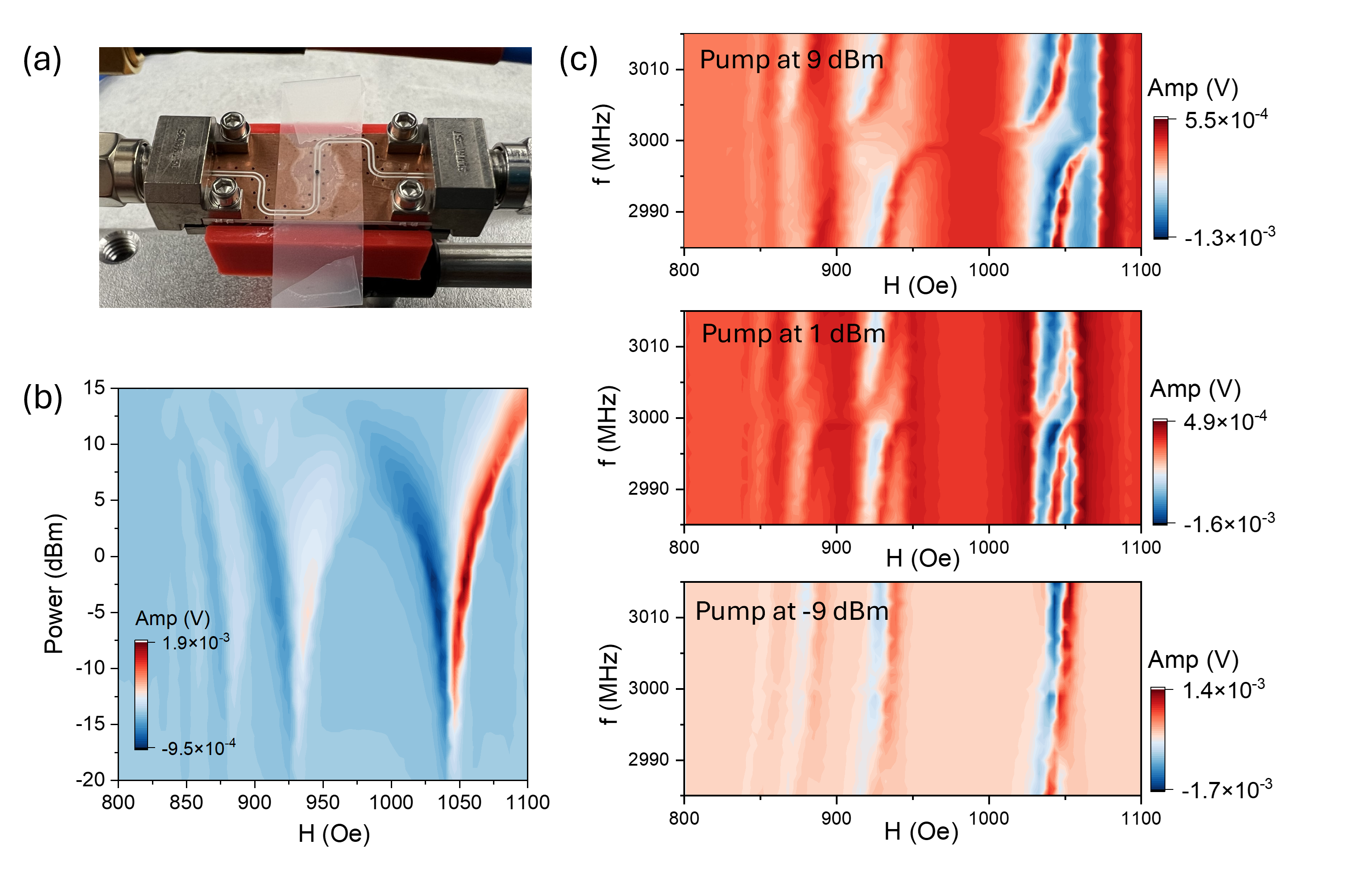}
 \caption{\textcolor{black}{Results for the 0.5-mm YIG sphere. (a) Photograph picture of the 0.5-mm YIG sphere on the CPW. (b) The power dependent spectrum of the magnetostatic modes. (c) The anti-crossing feature at selective power levels. Top: 9 dBm, middle: 1 dBm, and bottom: -9 dBm. }}
 \label{fig:small_sphere}
\end{figure}

\textcolor{black}{To further check the effect from the non-uniform CPW field, we also tested another YIG sphere with a smaller diameter, i.e., 0.5 mm nominal. The results are shown in Fig.\ref{fig:small_sphere}. All the key anti-crossing features remain the same for the small YIG sphere, however, only four Walker modes can be prominently observed. For the higher-order Walker modes, excitation becomes less efficient due to the smaller mode profiles compared to the same CPW antenna. }

\section{\\B: Identification of magnetostatic modes}

\begin{figure}[htb]
 \centering
 \includegraphics[width=4 in]{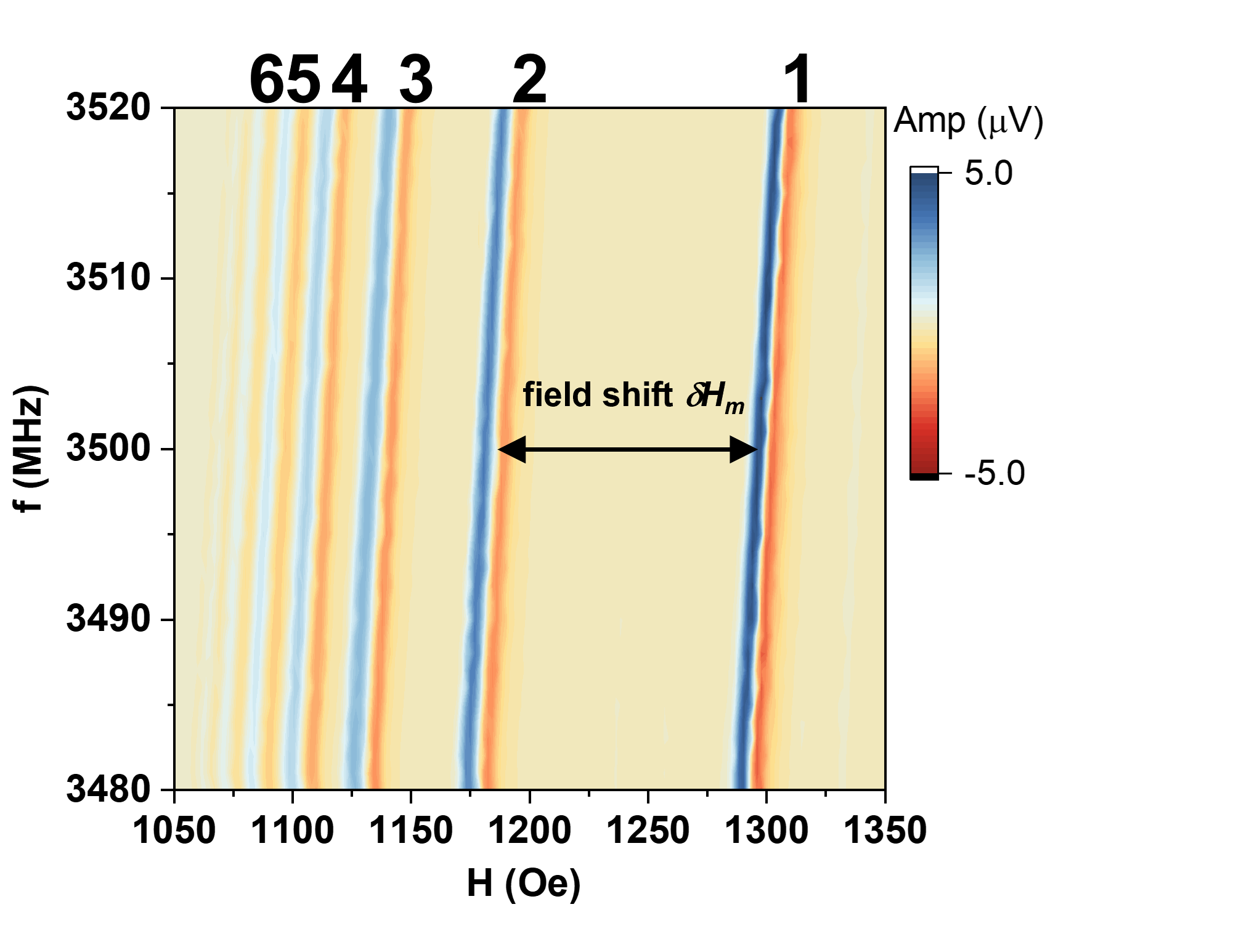}
 \caption{Linear dispersion of the magnetostatic modes near 3500 MHz (center frequency). }
 \label{fig:ms_mode}
\end{figure}

The properties of the magnetostatic modes are determined by measurement spectra under no pump and at a probe power of -10 dBm. An example 1-D spectrum measured at 3500 MHz is shown below. A total of 6 modes can be clearly identified and for all signal frequencies, Fig.\ref{fig:ms_mode}. All modes have linear dispersion with the magnetic field, $H$, with the same slope and are fitted to: $f_m(H)=\gamma (H + \tilde{\delta} H_m)$, where $\gamma$ is the gyromagnetic ratio for all modes and equals to 2.8 MHz / Oe, and $\tilde{\delta} H_m$ is the different field shifts. The 1-st mode, corresponding to $m$ = 1, is the ferromagnetic resonance (FMR) mode. The saturation magnetization from the fitting of the modes yields a value of $4\pi M_s=1711$ G, close to the standard value of $4\pi M_s=1750$ G. The obtained value of $4\pi M_s$ is rather robust, and very weakly depends on the fitting details. The excited sequence of modes is identified as $(m,m,0)$ sequence, where $(1,1,0)$ is FMR mode. These modes are circularly polarized. The precession amplitude of the mode does not depend on $z$ direction (uniform along the bias field). In the $(x, y)$ plane, precession amplitude scales as $\rho^{(m-1)}$ and precession phase is $(m-1)\phi$. The $m>1$ modes are non-uniform modes, and they can be excited efficiently when the CPW width is comparable to the YIG sphere radius, which is the present case. 
 
The field shift, $\tilde{\delta} H_m$, has contributions from the measurement system including but not limited to the crystallographic anisotropy and/or partial screening of the dipolar rf magnetic field. However, they have the same effect on all the spin wave modes, therefore, what is more important is the true field shift, $\delta H_m$, which is defined as the field shift of the $m$-th mode to the FMR mode, caused only by the deviation of the mode structure from the uniform one. Therefore, the mode dispersion can be written as: $f_m(H)=\gamma (H + H_c + \delta H_m)$, where $H_c$ denotes the uniform field shift from the system and $\delta H_m$ is the 'true' field shift of $m$-th mode due to solely the deviation of mode structure from uniform.  

\textcolor{black}{According to the magnetostatic theory for the  the $(m,m,0)$ Walker mode, the effective field for the $m$-th mode satisfies [1-5]:  
\begin{equation}
    \frac{f}{\Gamma 4\pi M_s}-\frac{H_{0,m,m}}{4\pi M_s} + \frac{1}{3}=\frac{m}{2m+1},
\label{Eq:H_mm}
\end{equation}
from Eq.\ref{Eq:H_mm}, we can calculate: $H_{0,1,1} = 4\pi M_s[\frac{f}{\Gamma 4\pi M_s}]$, and therefore the field shift, 
$\delta H_m = H_{0,m,m} - H_{0,1,1} = 4\pi M_s [\frac{f}{\Gamma 4\pi M_s}+\frac{1}{3}-\frac{m}{2m+1}] -  4\pi M_s[\frac{f}{\Gamma 4\pi M_s}] = 4\pi M_s [\frac{1}{3}-\frac{m}{2m+1}] = - 4\pi M_s [\frac{1}{3} \frac{m-1}{2m+1}]$,
in which we define the shift coefficient $\xi_m =$ $\frac{1}{3} \frac{m-1}{2m+1}$. As a result, theoretical true field shift $\delta H_m$, referencing to the FMR (1-st mode), can be expressed as $\delta H_m = \xi_m 4\pi M_s $, where $\xi_m= \frac{1}{3}\frac{m-1}{2m+1}$. The theoretical and experimental $\delta H_m$ are extracted and compared in Table below.  }

\begin{center}
\begin{tabular}{|p{1.5cm} p{1.7cm} p{1.5cm} p{1.5cm}|} 
 \hline
 $\textbf{m}$ & $\xi_m$ & $\delta H_m^\mathrm{exp}$ & $\delta H_m^\mathrm{theo}$  \\ [1 ex] 
 \hline\hline
 1 & 0 & 0 & 0  \\ 
 \hline
 2 & 0.06667 & 114.\underline{7} & 114.\underline{1} \\
 \hline
 3 & 0.09524 & 162.\underline{5} & 162.\underline{9}  \\
 \hline
 4 & 0.11111 & 18\underline{9.4} & 19\underline{0.1} \\
 \hline
 5 & 0.12121 & 207.\underline{2} & 207.\underline{4}  \\
 \hline
 6 & 0.12821 & 219.\underline{6} & 219.\underline{4}  \\ [1ex] 
 \hline
\end{tabular}
\end{center}

\section{\\C: Three-magnon scattering threshold fields and frequencies }

In three magnon process, the parametric excitation by mode $m$ is possible when: $f_m(H)>2f_{min}(H)$, where $f_{min}(H)=\gamma(H + H_c - \frac{1}{3} 4\pi M_s)$ is the minimum magnon frequency in a magnetic sphere. This then determines the maximum field $H_{m,crit}$ and the maximum frequency $f_{m,crit}$, for which three-magnon decay of mode $m$ is possible:

\begin{equation}
    H_{m,crit} = (\frac{2}{3} + \xi_m) 4 \pi M_s - H_c,
\label{Eq:H_crit}
\end{equation}

\begin{equation}
    f_{m,crit} = \gamma (\frac{2}{3} + 2\xi_m) 4 \pi M_s + f_{offset},
\label{Eq:f_crit}
\end{equation}

The calculated $H_{m,crit}$ and $f_{m,crit}$ are summarized in Table below. 

\begin{center}
\begin{tabular}{|p{2.5cm} p{2.5cm} p{2.5cm}|} 
 \hline
 $\textbf{m}$ & $H_{m,crit}$(Oe) & $f_{m,crit}$(MHz)  \\ [1 ex] 
 \hline\hline
 1 & 1191 & 3193 \\
 \hline
 2 & 1305 & 3832  \\
 \hline
 3 & 1354 & 4106 \\
 \hline
 4 & 1381 & 4258  \\
 \hline
 5 & 1398 & 4354  \\
 \hline
 6 & 1410 & 4421  \\ [1ex] 
 \hline
\end{tabular}
\end{center}

\begin{figure}[htb]
 \centering
 \includegraphics[width=3.3 in]{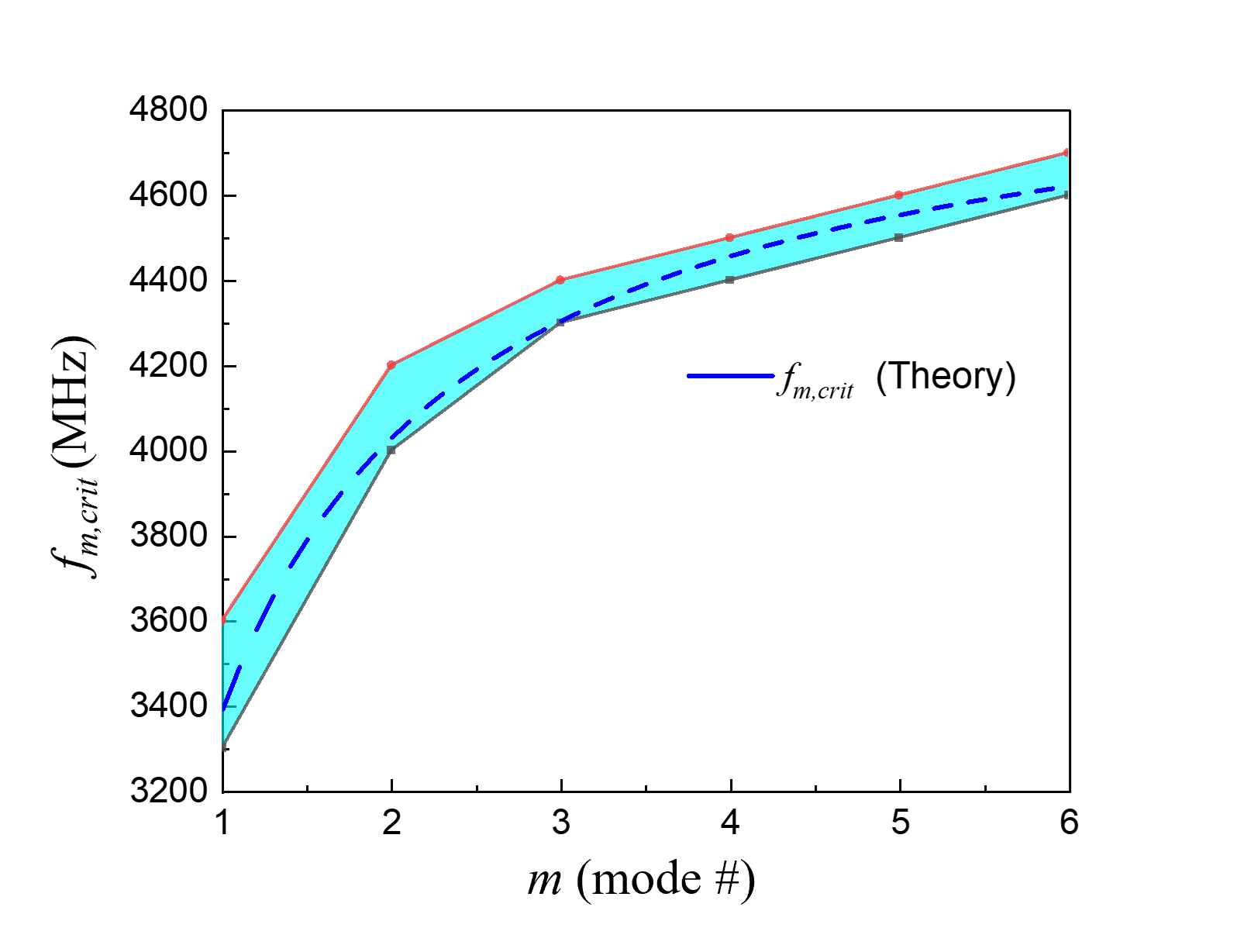}
 \caption{Calculated (dashed line) and experimental (shaded area) $f_{m,crit}$ values.}
 \label{fig:fcrit}
\end{figure}

\textcolor{black}{The $H_{m,crit}$ and $f_{m,crit}$ are sensitive to experimental fitting parameters, e.g., $M_s$, $\gamma$, and $H_c$. They may shift by $\sim 100$ Oe and $\sim 200$ MHz ($f_{offset}$), respectively, but qualitative picture will be the same. For example, a strong enough pumping may parametrically excite magnons, which are close to, but not exactly at the resonance line. also, the four-magnon nonlinearity (ignored here) may also shift all dispersion relations. We show in Fig.\ref{fig:fcrit} the calculated the $f_{m,crit}$ with a $f_{offset}=200$ MHz which nicely agrees with the experimental data. The experimental values were extracted and shown in the same plot.  }\textcolor{black}{Our experimental analysis on the index-dependent frequency thresholds of the three magnon scattering nonlinearity (of the Walker modes) and the congruent anticrossing gap agrees with three magnon scattering characteristics. For frequencies above the frequency threshold of a certain mode, the mode linewidth increases weakly as a function of the drive power. Since 3-magnon scattering is forbidden, the 4-magnon scattering as the remaining mechanism, is weak up to the maximum pumping power used in the present work. In addition, the 4-magnon scattering is a high order effect, compared to the three magnon scattering, that mostly dominates in large angle rotations, e.g. switching, but is not dominant in the small angle precession as of ferromagnetic resonance.}

\begin{figure*}[htb]
 \centering
 \includegraphics[width=7 in]{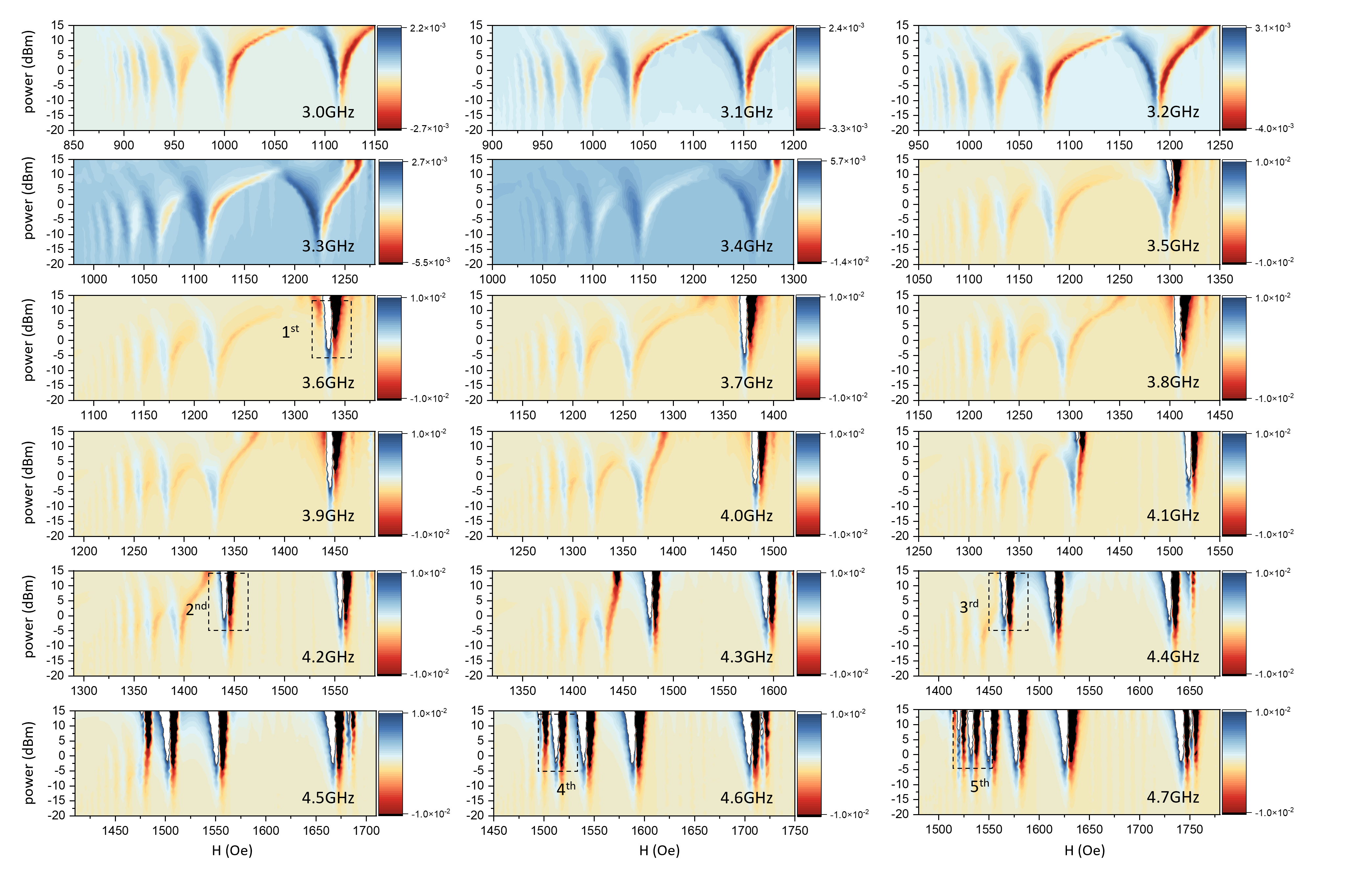}
 \caption{Measured power dependent spectra of the magnetostatic modes. The three-magnon cutoff frequency of each mode can be identified.}
 \label{fig:all_power}
\end{figure*}

In Fig. \ref{fig:all_power}, we show the comprehensive power dependent spectra, measured for a range of frequencies (3.0 – 4.7 GHz, at a step of 0.1 GHz) to elucidate the three-magnon process and to identify the cutoff threshold. The frequency threshold for the modes is identified as: 1st (3.6 GHz), 2nd (4.2 GHz), 3rd (4.4 GHz), 4th (4.6 GHz), and 5th and higher (4.7 GHz).

\section{\\D: Anti-crossing gap at different pump power levels}

\begin{figure*}[htb]
 \centering
 \includegraphics[width=0.8\columnwidth]{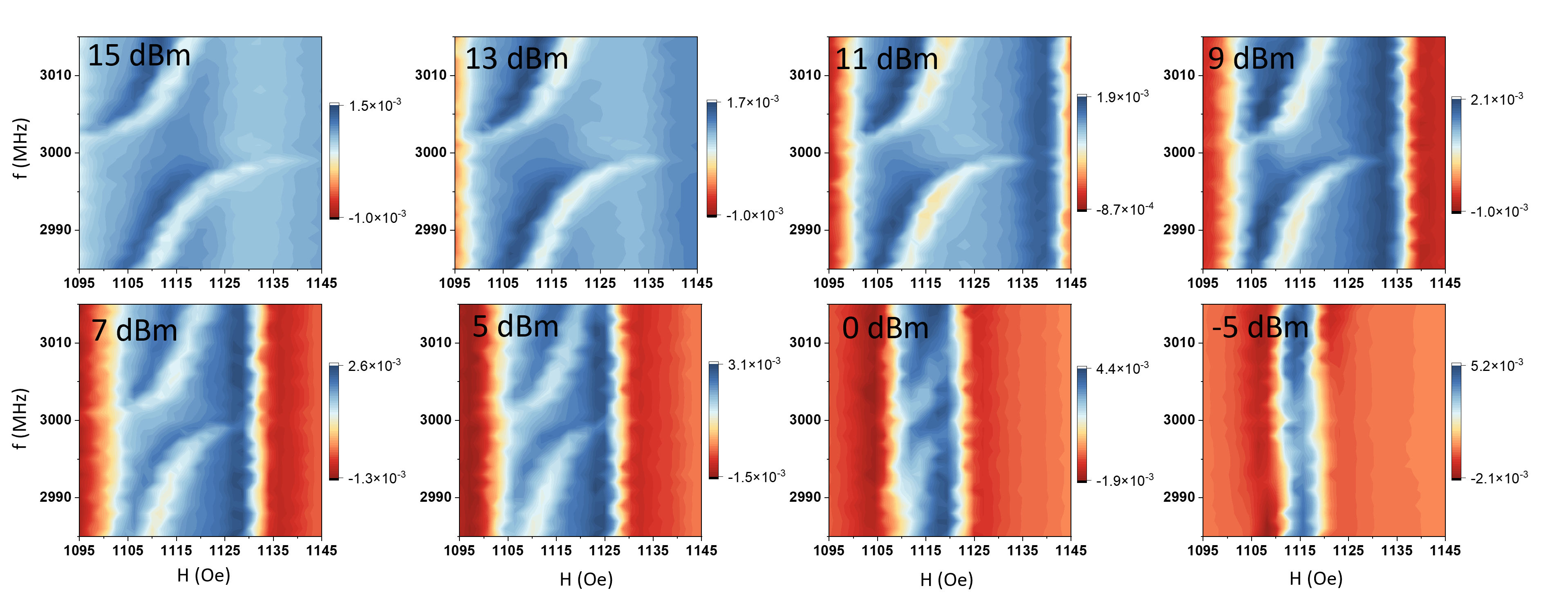}
 \caption{The zoom-in plots of the observed anti-crossing feature for the FMR mode at different power levels (15, 13, 11, 9, 7, 5, 0, -5 dBm) under a pump frequency at 3.0 GHz (field centered around 1110 Oe).}
 \label{fig:power_dpdce}
\end{figure*}

Figure \ref{fig:power_dpdce} shows the zoom-in plots of the observed anti-crossing 'gap' feature for the FMR mode at different power levels under a pump at 3.0 GHz (field centered around 1110 Oe). Upon power decreasing, the condensate 'range' shrinks, and the anti-crossing gap size also reduces. No apparent anti-crossing feature can be recognized when power level is below 0 dBm. \textcolor{black}{Due to our field-modulation technique (with modulation coils and lock-in amplifier detection), no parasitic microwave background exists, thus no background subtraction is performed for the raw data of our experiment, in contrast to the vector network analyzer technique.}
\textcolor{black}{The non-uniformity of the ac field allows us to excite and examine the physical behaviors of the higher-order Walker modes and conclude the frequency threshold relating to the three-magnon process and the anti-crossing gap. However, it does not affect the physical behavior itself (anticrossing gap and power dependence), as all the excited modes share the same spectra and characteristics, and thus can be generalized to different sizes of YIG spheres.}

\section{\\E: Application of double pumps  }

\begin{figure*}[htb]
 \centering
 \includegraphics[width=0.7\columnwidth]{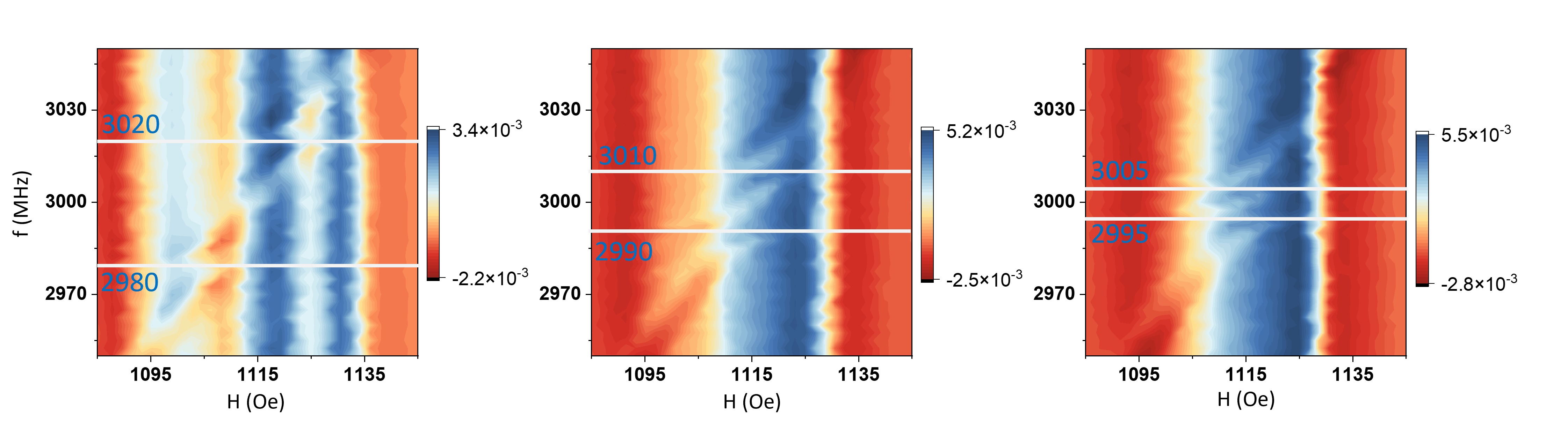}
 \caption{Measured spectra with two pump signals applied simultaneously. Power is fixed at 10 dBm. Left: pumps at 2980 / 3020 MHz; Middle: pumps at 2990 / 3010 MHz; right: pumps at 2995 / 3005 MHz. }
 \label{fig:double}
\end{figure*}

To demonstrate the generality of the pump induced effects, we also tried to introduce two pumps simultaneously that are close to each other and measured the dispersion spectra. The result is shown in Fig. \ref{fig:double}. The powers of the two pumps are the same and both equal to 10 dBm. Three examples are shown. Left: pumps at 2980 and 3020 MHz; Middle: pumps at 2990 and 3010 MHz; right: pumps at 2995 and 3005 MHz. Two corresponding anti-crossing gaps can be clearly identified right at the pump signals. 

\textcolor{black}{Applying a double pump (with close but different frequencies) results in an overlap of the two condensates centered at the two respective pumps. The enhanced condensate amplitude will increase each anti-crossing gap at the two pumps, which will ultimately merge into one larger gap when the two pumps are extremely close to each other, as in the case of doubling the pump power.}

\section{\\F: Derivation of magnon dispersion and nonlinear coupling strength in three-magnon scattering}

\begin{figure}[htb]
 \centering
 \includegraphics[width=0.4\columnwidth]{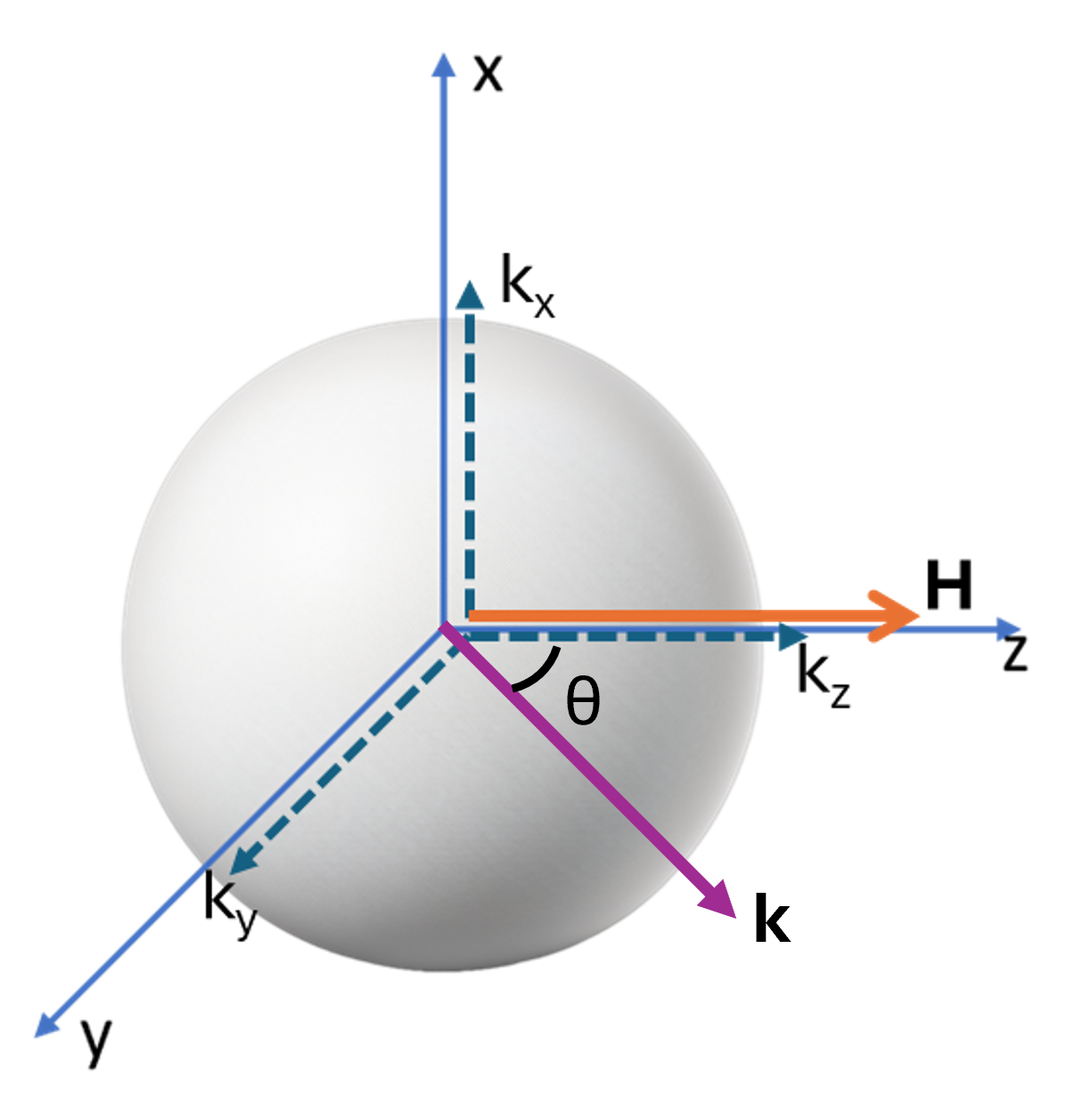}
 \caption{Coordinates of the sphere system, the $z$ axis is defined by the applied field $\textbf{H}$, $\theta$ is the angle between wavevector $\kbf$ and the applied field $\textbf{H}$.}
 \label{fig:coord}
\end{figure}

We rewrite the Landau–Lifshitz equation in the \textbf{k} space in the framework of magnon with counterclockwise polarization $a_k$ and clockwise $a_k^-$, 
\begin{eqnarray}
\dot{m}^{+}=\sum\dot{a}_k e^{\imath k r}=\imath m^+[\omega_H+\omega_{dz}+\omega_{ex}l^2\nabla^2m_z]\\
-\imath m_z[\omega_{s}e^{\imath \omega t}+\omega_{ex}l^2\nabla^2m^{+}+\omega_{dx}+\imath \omega_{dy}] \\
m^+=m_x+\imath m_y=\sum a_k e^{\imath k r} \\
m^-=m_x-\imath m_y=\sum a_{k}^- e^{\imath k r} \\
a_k^-=a_{-k}^*~~~\textrm{(conjugate relation)}  \\
m_z=1-\frac{1}{2}\sum_{k k'}a_{k'}a_{k'-k}^*e^{\imath k r}  
\end{eqnarray}
where the $z$ coordinate is defined by the direction of the applied magnetic field H, as shown in Fig.\ref{fig:coord}. The $m_x$, $m_y$, and $m_z$ are the components of the magnetization vector $\mathbf{m}$ in the real space. $\omega_H$ and $\omega_{ex}$ are the constant angular frequency corresponding to the Zeeman energy and exchange energy respectively. $l$ is the constant exchange length. $\omega_{dx}$, $\omega_{dy}$ and $\omega_{dz}$ are the angular frequencies, induced by the demagnetization field $\mathbf{H}_{d}$, which is dependent on the wavevector $\kbf$,  $\mathbf{H_d}=-\sum_\kbf \frac{\kbf(\kbf\cdot\mathbf{M_k})}{k^2}$. 

For one specific \textbf{k} $\neq0$,  at the right side, $m^+$ has a significant amplitude when mapping to $e^{\imath k r}$, thus the first order is 
\begin{equation}
[\omega_H+\omega_{dz}+\omega_{ex}l^2\nabla^2m_z]=(\omega_H-N_z\omega_m)a_k
\end{equation} 
Where $\nabla^2m_z\sim0$ since this $z$ component of magnetization is pinned by the external field. While $m_z$ is considered as constant that approximate $M_{kz}\sim0$, the first order with $e^{\imath k r}$ originates from the field component at  wavevector $\kbf$ 
\begin{eqnarray}
\omega_{ex}l^2\nabla^2m^{+}\to -k^2\omega_{ex}l^2 a_k \\
\omega_{dx}\to -4\pi \frac{k_x \bm{k\cdot M_k} }{k^2} \to -4\pi \frac{k_x^2 M_{kx}+k_x k_y M_{ky} }{k^2}\\ 
\omega_{dy}\to -4\pi \frac{k_y \bm{k\cdot M_k} }{k^2} \to -4\pi \frac{k_x k_y M_{kx}+k_y^2 M_{ky} }{k^2} 
\end{eqnarray} 
Rewrite $\omega_{dx}$ and $\omega_{dy}$ in $a_k$ and $a_{-k}^*$, 
\begin{eqnarray}
\omega_{dx}+\imath \omega_{dy}\to -4\pi \frac{k_x^2+k_y^2}{2k^2}a_k-4\pi\frac{(k_x+\imath k_y)^2}{2k^2}a_{-k}^*\\
\omega_{dx}+\imath \omega_{dy} \to -\omega_M \frac{|k_T|^2}{2k^2}a_k-\omega_M\frac{k_T^2}{2k^2}a_{-k}^*
\end{eqnarray} 
where $k_T=k_x+\imath k_y$. 
Summing up in the first order 
\begin{eqnarray}
-\imath\dot{a}_k=A_ka_k+B_ka_{-k}^* \\
A_k=\omega_H-\omega_MN_z+\omega_{ex}l^2k^2+\omega_M \frac{|k_T|^2}{2k^2} \\
B_k=\omega_M \frac{k_T^2}{2k^2} \\
\end{eqnarray}

Deriving $\dot{a}_k^-$ and write it in $a_{-k}^*$
\begin{eqnarray}
\dot{m}^{-}=\sum\dot{a}_k^- e^{\imath k r}=-\imath m^-[\omega_H+\omega_{dz}+\omega_{ex}l^2\nabla^2m_z]\\
+\imath m_z[\omega_{s}e^{\imath \omega t}+\omega_{ex}l^2\nabla^2m^{-}+\omega_{dx}-\imath \omega_{dy}]
\end{eqnarray}
\textcolor{black}{The physical feature of the magnon is right-handed, the same as the case in ferromagnets. However, in the mathematical treatment, we multiply a negative sign to the magnetization component as of $m_x-\imath m_y$, producing an illusion of a left-handed magnon existing in our system.}
\subsection{\textcolor{black}{Magnon Dispersion}}
\textcolor{black}{Using similar steps, for the first order of the first term on the right side 
\begin{eqnarray}
-\imath m^-[\omega_H+\omega_{dz}]\to-\imath a_k^-(\omega_H-N_z\omega_m) 
\end{eqnarray}
Similarly, the first order of the second term in the right side: 
\begin{eqnarray}
\omega_{ex}l^2\nabla^2m^{-}\to -k^2\omega_{ex}l^2 a_k^- \\
\omega_{dx}-\imath \omega_{dy}\to -4\pi \frac{k_x^2+k_y^2}{2k^2}a_k^{-}-4\pi\frac{(k_x-\imath k_y)^2}{2k^2}a_{k}
\end{eqnarray}}

\textcolor{black}{As $a_k^-=a_{-k}^*$, 
\begin{eqnarray}
\imath \dot{a}_{-k}^*= A_k a_{-k}^{*}+B_k^* a_k
\end{eqnarray}}

\begin{figure}[htb]
 \centering
 \includegraphics[width=0.7\columnwidth]{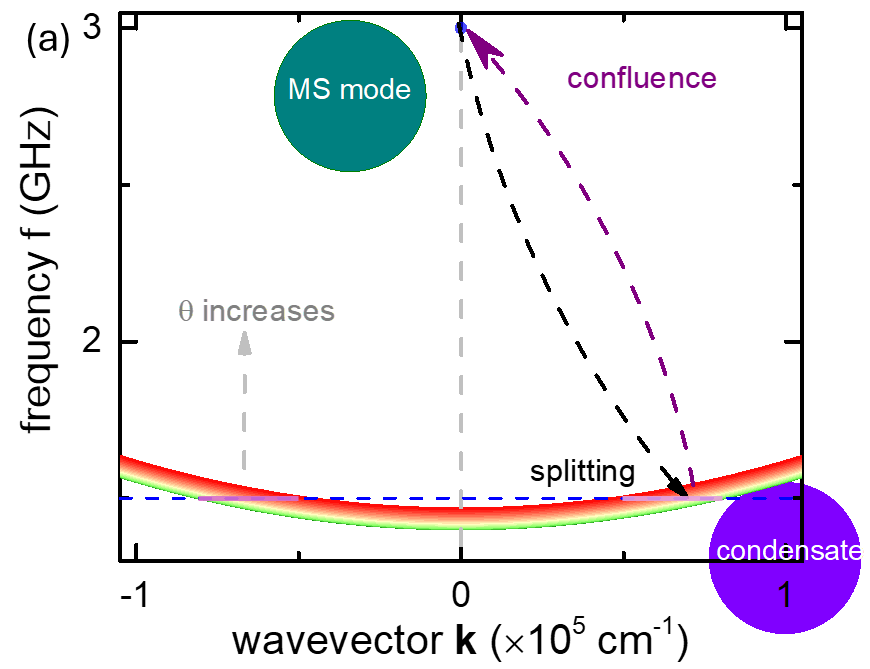}
 \caption{\textcolor{black}{(a) The magnon dispersion as of the intrinsic frequency versus the wavevector $\kbf$ along varying directions, defined by the angle $\theta$ between the wavevector $\kbf$ and the applied field $\textbf{H}$. At the same $|\kbf|$, the frequency monotonically increases with the increasing angle between $0$ to $\pi/2$, due to the component $\sin^2(\theta)$ in Eq.\ref{eq:dispersion}. The frequency is degenerate for the wavevector $k$ at $\theta$ and $\pi-\theta$. The figure takes the three magnon scattering for the first magnetostatic mode $m$ =1 at frequency $f_d$=3.0 GHz, as an example. The dashed crossline is at $\frac{f_d}{2}$= 1.5 GHz, indicating the three magnon scattering occurs between one $\kbf$=0 magnon mode and multiple degenerate $\kbf\neq0$ modes ($|\kbf|$,$\theta$,$\phi$) with intrinsic frequency at $\frac{f_d}{2}$, highlighted in the blue region.  }}
 \label{fig:dispersion}
\end{figure}

\textcolor{black}{The first order depicts the magnon spectrum: the intrinsic angular frequency of the magnon at wavevector $\kbf$, is summarized in Eq. \ref{eq:dak_dt}
\begin{equation}
\begin{pmatrix} 
\dot{a}_k  \\
\dot{a}_{-k}^*  
\end{pmatrix}=
\begin{pmatrix} 
\imath A_k & \imath B_k \\
-\imath B_k^* & -\imath A_k 
\end{pmatrix}
\begin{pmatrix} 
a_k \\
a_{-k}^*  
\end{pmatrix}
\label{eq:dak_dt}
\end{equation}}
\textcolor{black}{where 
\begin{eqnarray}
A_k=\omega_H-\omega_MN_z+\omega_{ex}l^2k^2+\omega_M \frac{|k_T|^2}{2k^2} \\
B_k=\omega_M \frac{k_T^2}{2k^2} 
\end{eqnarray}}
\textcolor{black}{Diagonalizing the matrix yields the magnon dispersion of the magnon energy $\hbar\omega_k$ versus the wavevector $\kbf$,
\begin{eqnarray}
\omega_k&=&\sqrt{A_k^2-|B_k|^2} \nonumber \\
&=&\sqrt{(\omega_H-\omega_MN_z+\omega_{ex}l^2k^2+\omega_M \frac{|k_T|^2}{k^2})\times(\omega_H-\omega_MN_z+\omega_{ex}l^2k^2)} \label{eq:dispersion}
\end{eqnarray}}
\textcolor{black}{where $\gamma=1.76\times10^7$ s$^{-1}$ Oe$^{-1}$, $4\pi M_s=1711$ Oe, $N_z=4\pi/3$ and $\omega_{ex}l^2=5.4\times10^{-9}$ Oe cm$^2$. At the applied field $\textbf{H}=1071$ Oe corresponding to the resonance frequency of 3 GHz from the Kittel equation, the magnon dispersion depicting the magnon frequency $f_m$ versus the wavevector $\kbf$ is in Fig. ~\ref{fig:dispersion}. At a fixed magnetostatic frequency $f_d$, the three magnon scattering occurs in an available k-band, which has a width in $\kbf$ but the frequency of this band is well-defined $f_d$/2, half of the magnetostatic frequency due to energy conservation, as shown in Fig. \ref{fig:dispersion}.} 

\subsection{Coupling Strength in Three Magnon Scattering}
We add the second order term, which determines the coupling between the magnons $\kbf\neq0$ and $\kbf=0$. This second-order term comes from $\omega_{dz}*m^+$. while with $\kbf-\mathbf{k'}$ term in $m^+$, the corresponding term  $\mathbf{k'}$ in $\omega_{dz}$ to yield $\kbf$ component is 
\begin{eqnarray}
\omega_{dz}&=&\sum_k -4\pi\gamma\frac{k_z(\kbf \cdot \mathbf{M_k})}{k^2}\nonumber\\
&=&-\omega_M\frac{k_z'(\mathbf{k'} \cdot \mathbf{m_{k'}})}{k'^2} \nonumber\\
&=&-\omega_M\frac{k_z'}{k'^2}(k_x'm_{k'}^x+(k_y'm_{k'}^y) \nonumber \\
&=&-\frac{\omega_M}{2}(k_{T'}{a}_{-k'}^*+k_{T'}^{*}{a}_{k'})\frac{k_z'}{k'^2}
\label{eq:3ms_coupling}
\end{eqnarray}
Thus the $\kbf$ component  of $m^+\omega_{dz}$ produce the three magnon coupling effect is $\sum_{k'\neq0}-a_{k-k'} \frac{\omega_M}{2}(k_{T'}{a}_{-k'}^*+k_{T'}^{*}{a}_{k'})\frac{k_z'}{k'^2}$. Following  law of momentum conservation $\mathbf{k'}=\kbf$,  the coupling term is simplified to $-a_{0} \frac{\omega_M}{2}(k_{T}{a}_{-k}^*+k_{T}^{*}{a}_{k})\frac{k_z}{k^2}$.

Include this coupling term between the $\kbf=0$ and $\kbf\neq0$ magnons in Eq. \ref{eq:dak_dt}, this yields the dynamics equation with the three magnon coupling in a sphere in Eq. \ref{eq:dak_dt_3MS}
\begin{equation}
\begin{pmatrix} 
\dot{a}_k  \\
\dot{a}_{-k}^*  
\end{pmatrix}=
\begin{pmatrix} 
\imath A_k-\imath \omega_M a_0k_T^* \frac{k_z}{2k^2} & \imath B_k -\imath \omega_M \frac{k_zk_T}{2k^2}a_0\\
-\imath B_k^*+\imath\omega_M \frac{k_zk_T^*}{2k^2}a_0^* & -\imath A_k +\imath \omega_M a_0^*k_T \frac{k_z}{2k^2}
\end{pmatrix}
\begin{pmatrix} 
a_k \\
a_{-k}^*  
\end{pmatrix}
\label{eq:dak_dt_3MS}
\end{equation}

\textcolor{black}{Thus the coupling strength $V$ is $\hbar \omega_M k_T \frac{k_z}{2k^2}$, set the angle between $\kbf$ and $z$ axis as $\theta$, the coupling strength is rewritten as $V=\hbar \omega_M \frac{\sin(2\theta)}{4} $. When $\theta=\pi/4$, it has the maximum coupling strength. }

\section{\\G: Time-domain measurements}

\begin{figure}[htb]
 \centering
 \includegraphics[width=1.0\columnwidth]{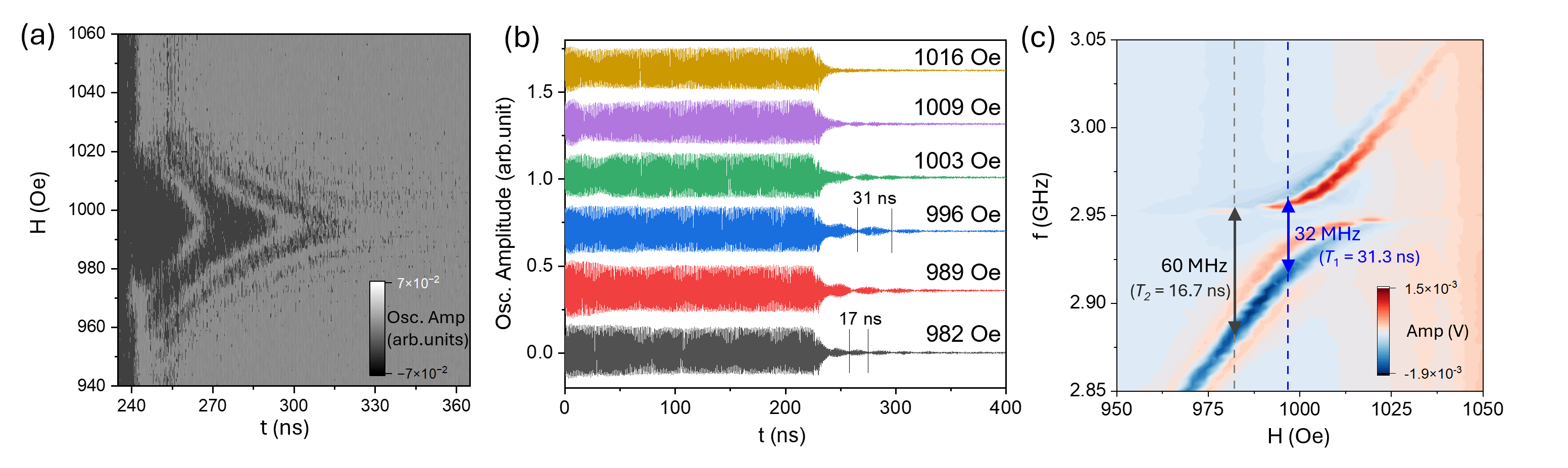}
 \caption{\textcolor{black}{(a) Contour plot of the Rabi-like oscillation time traces showing the magnetic-field dependent splitting--confluence oscillation dynamics, measured for the 0.5-mm YIG sphere sample at a pump power of 15 dBm. (b) 1-D time traces at selective magnetic field values near the coupling regime (pump power at 15 dBm). (c) The corresponding frequency-domain measurement under the same pump (15 dBm) and with an additional probe (power at $-10$ dBm).} }
 \label{fig:TD}
\end{figure}

\textcolor{black}{To verify that the three-magnon splitting and confluence indeed manifests as a Rabi-like process, we performed additional time-domain measurements following the method outlined in our earlier experiment [6], and the result is summarized in Fig.\ref{fig:TD}. As can be seen in Fig.\ref{fig:TD}(a), the splitting--confluence dynamics exhibit a magnetic-field dependence. The Rabi-like oscillation period can be estimated from the time traces, in Fig.\ref{fig:TD}(b); for example, 31 ns at $H = 996$ Oe and 17 ns at $H=982$ Oe. These values are in good agreement with the time period calculated from the anti-crossing gap in the spectral measurement, in Fig. \ref{fig:TD}(c), i.e., $T_1$=1/(32-MHz) = 31.3 ns for $H = 996$ Oe (close to the center of the gap), and $T_2$=1/(60-MHz) = 16.7 ns for $H = 982$ Oe (away from the gap), respectively.  }

\newpage 

\textbf{References:} \newline 

[1] P. Roschmann and H. Dotsch, physica status solidi (b) 82, 11 (1977).

[2] P. Fletcher and R. Bell, Journal of applied physics 30, 687 (1959).

[3] R. L. White, Journal of Applied Physics 31, S86 (1960).

[4] L. R. Walker, Physical Review 105, 390 (1957).

[5] A. Leo, A. G. Monteduro, S. Rizzato, L. Martina, and G. Maruccio, Physical Review B 101, 014439 (2020).

[6] T. Qu, A. Hamill, R. H. Victora, and P. A. Crowell, Phys. Rev. B 107, L060401 (2023).

\end{document}